\documentclass[reprint,aps,prb,superscriptaddress]{revtex4-2}
\usepackage{amsmath,amssymb,bm,mathtools}
\usepackage{graphicx}
\usepackage[colorlinks=true, allcolors=blue]{hyperref}

\newcommand{\yW}{y_{\min}}
\newcommand{\aW}{a_{\mathrm{W}}}
\newcommand{\kWS}{\kappa_{\mathrm{WS}}}
\newcommand{\Es}{E_{\mathrm{s}}}
\newcommand{\cW}{D}
\newcommand{\Wmin}{W_{\min}}

\begin{document}
\title{Noncommutative Skin Effect from Commuting Coordinates}
\author{Zheng Wei}
\affiliation{Center for Advanced Quantum Studies, School of Physics and Astronomy, Beijing Normal University, Beijing 100875, China}
\affiliation{Key Laboratory of Multiscale Spin Physics (Ministry of Education), Beijing Normal University, Beijing 100875, China}
\author{Su-Peng Kou}
\email{spkou@bnu.edu.cn}
\affiliation{Center for Advanced Quantum Studies, School of Physics and Astronomy, Beijing Normal University, Beijing 100875, China}
\affiliation{Key Laboratory of Multiscale Spin Physics (Ministry of Education), Beijing Normal University, Beijing 100875, China}

\begin{abstract}
Noncommuting coordinates recur in Landau levels, string theory, and quantum gravity, whereas the coordinates of real systems commute. Projected onto a lattice band, two coordinates no longer commute; their commutator is given by the Berry curvature. When we apply a gain-loss gradient across a lattice strip, this noncommutativity makes bulk-band eigenmodes accumulate at one end of the strip, producing a noncommutative skin effect. For a uniform gradient, we obtain an analytic formula: the inverse skin length is proportional to the gradient and to the Berry-curvature flux of all states below the eigenmode energy, and inversely proportional to the summed speeds of the right- and left-moving eigenmodes. Bands with zero Chern number also show the effect when this flux is nonzero. For weak gradients, a smooth confining potential across the strip sets the eigenmode positions. For a single-minimum potential symmetric about its minimum, the skin effect is suppressed for eigenmode energies below the separatrix energy but not above it, irrespective of the Chern number. Larger gradients restore the formula for both open boundaries and smooth confinement across the strip. In circuits, grounding elements provide the gradient and the potential; the inverse skin length can be read from the measured impedance matrix. Noncommutative space need not be engineered: a band with Berry curvature is itself a noncommutative space, and a band on which this flux is nonzero is a platform for the noncommutative skin effect.
\end{abstract}

\maketitle

\section{Introduction}\label{sec:intro}

Two spatial coordinate operators that do not commute, $[\hat\xi,\hat\eta]=i\theta$ (with $\theta$ a constant of dimension length squared), appear again and again in physics: as the guiding-center coordinates of the lowest Landau level in a strong magnetic field~\cite{Bellissard1994,Ezawa2013}, on D-branes in string theory~\cite{Seiberg1999}, and in proposals for a minimal length in quantum gravity~\cite{Doplicher1995,Douglas2001}. Noncommutative quantum mechanics starts from such a space and studies quantum mechanics in it~\cite{Gamboa2001}. In real systems, however, the two spatial coordinates themselves commute: even in a magnetic field, the position operators become the noncommuting guiding-center coordinates only after a strong field has confined the motion to a single Landau level. This raises a basic question: in a real system whose coordinates commute, how can the physics of noncommutative space arise?

The lattice itself offers an answer that needs no magnetic field. The two real-space coordinate operators commute, but after projection onto an isolated band, the commutator of the two band-projected coordinates is given by the Berry curvature of that band~\cite{XiaoChangNiu2010}; wherever the Berry curvature is nonzero, the two projected coordinates no longer commute. The algebra of projected density operators on a band with nonzero Chern number is isomorphic to that of the lowest Landau level at long wavelengths when the Berry curvature is uniform~\cite{Parameswaran2012}. In Hermitian systems the consequences of this noncommutativity are familiar: Bloch electrons acquire an anomalous velocity, and the system shows a Hall response~\cite{XiaoChangNiu2010}. The quantum geometric tensor and the anomalous Hall drift have also been measured~\cite{Gianfrate2020}. What the same noncommutativity brings to non-Hermitian systems is an open question.

A hallmark of non-Hermitian systems is the non-Hermitian skin effect: under open boundary conditions (OBCs), a macroscopic number of eigenstates accumulate exponentially at the boundary, and the spectrum collapses accordingly~\cite{Yao2018,Yokomizo2019}. In translation-invariant lattices it originates from point-gap topology~\cite{Okuma2020}. It invalidates the bulk-boundary correspondence of Bloch bands~\cite{Yao2018,zhang_CorrespondenceWindingNumbers_2020,borgnia_NonHermitianBoundaryModes_2020,yang_NonHermitianBulkBoundaryCorrespondence_2020,kawabata_SymmetryTopologyNonHermitian_2019}, has given rise to non-Bloch band theory~\cite{Yokomizo2019,wang_AmoebaFormulationNonBloch_2024}, and has been observed in photonic, electric-circuit, acoustic, cold-atom, and quantum-walk platforms~\cite{weidemann_TopologicalFunnelingLight_2020,helbig_GeneralizedBulkBoundary_2020,zhang_AcousticNonHermitianSkin_2021,Liang2022,Zhao2025,xiao_NonHermitianBulkBoundary_2020} (for reviews, see~\cite{ashida_NonHermitianPhysics_2020,Ding2022review,Okuma2023review}). Its variants and related phenomena include the critical skin effect~\cite{li_CriticalNonHermitianSkin_2020}, scale-free localization~\cite{li_ImpurityInducedScalefree_2021,li_ScalefreeLocalizationPT_2023}, the higher-order skin effect~\cite{zhang_HigherOrderNHSE_2021}, the geometry-dependent skin effect~\cite{zhang_UniversalNonHermitianSkin_2022,Wang2023GDSE,Zhou2023GDSE}, the boundary-insensitive skin effect~\cite{Wei2025}, and the reversal of the skin-effect direction~\cite{Li2022reversal,Yuce2024bipolar,Yang2026transverseSwitch}. It is driven not only by asymmetric hopping~\cite{Hatano1996} but also by gain and loss~\cite{yi_NonHermitianSkinModes_2020,jiang_TunableNonHermitianSkin_2024}, dissipative coupling~\cite{Huang2024dissipative}, Floquet driving~\cite{Lin2024Floquet,Sun2024Floquet}, sample geometry~\cite{Wang2023GDSE}, and spatially inhomogeneous imaginary potentials~\cite{Wei2025}. In two and higher dimensions, both the skin effect and non-Bloch band theory have been generalized~\cite{zhang_UniversalNonHermitianSkin_2022,wang_AmoebaFormulationNonBloch_2024}. Gain and loss along one direction, combined with a magnetic flux or with a band of nonzero Chern number, can also produce a skin effect along another direction. Dissipative flux ladders~\cite{Wu2022}, Hall ladders in synthetic dimensions~\cite{Ye2025}, and the chiral edge modes of Chern insulators with gain and loss~\cite{ZhuGong2022,Li2022edge,Ma2024} are examples.

On the noncommutative plane, where the commutator of the two coordinates is the constant $i\theta$, gain and loss along one coordinate likewise produce a skin effect along the other~\cite{WeiKouCPL}. Since this skin effect is caused by the noncommutativity of the coordinates, it is called the noncommutative skin effect. In a lattice, with $x$ and $y$ the two real-space directions and $(k,p)$ the conjugate quasimomenta, the commutator is given by the Berry curvature, which varies with $(k,p)$. Its integral over a whole band equals $2\pi$ times the Chern number, which takes integer values. On a band with zero Chern number, the Berry curvature itself can still be nonzero. A lattice sample must also be bounded in $y$, either by OBCs or by a smooth confining potential. We are thus led to ask: can gain and loss along $y$ produce a skin effect along $x$ in the bulk bands of a lattice, through the noncommutativity of band-projected coordinates? If so, which part of the commutator determines the skin effect, and what role does the boundary condition in $y$ play?

\begin{figure}[t]
\includegraphics[width=\columnwidth]{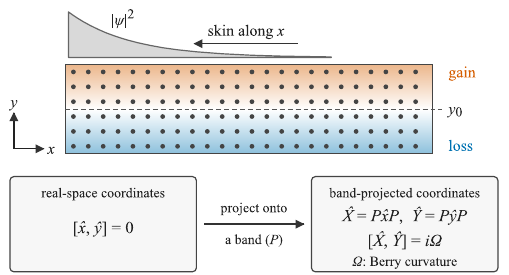}
\caption{Gain and loss along $y$ produce a skin effect along $x$ through the noncommutativity of band-projected coordinates. Top: a strip, long in $x$ and finite in $y$, with on-site gain and loss varying linearly in $y$; they vanish on the center line $y_0$, with gain above it and loss below. The gray envelope is the intensity $|\psi|^2$ of an OBC eigenmode, which accumulates at the small-$x$ end (as in Fig.~\ref{fig:modes}(a)). Bottom: the real-space coordinate operators commute; after projection onto a band with the projector $P$, $\hat X=P\hat xP$ and $\hat Y=P\hat yP$ do not commute, and their commutator is $i\Omega$, with $\Omega$ the Berry curvature [Eq.~\eqref{eq:Omega}].}
\label{fig:schematic}
\end{figure}

We answer these questions on a lattice strip that is translation invariant in $x$ and has a finite width in $y$ (Fig.~\ref{fig:schematic}). We project onto an isolated band, which we call the selected band. At fixed $k$, we refer to the energy levels of the motion along $y$ as the $y$ levels. Gain and loss varying linearly in $y$ then turn the $y$ levels into a ladder equally spaced along the imaginary axis. The real part of the eigenvalue of each level is its energy and the imaginary part is its gain rate. The gain rate is given by the hybrid Wannier center. The Wannier center moves with $k$, and its rate of change is precisely the average of the commutator over $p$. On one level of the ladder, there is one right-moving and one left-moving eigenmode at the same energy. The two have different $k$, hence different Wannier centers and different gain rates, so the eigenmode under OBCs in $x$ accumulates at one end. When the difference in gain rate between the two sides of the strip is much larger than the bandwidth of the selected band, the ladder forms within the strip. If, in addition, the gradient is weak enough not to mix in other bands, we obtain an analytic formula for the inverse skin length (Sec.~\ref{sec:ladder}). The inverse skin length is proportional to the gradient, proportional to the Berry-curvature flux of all states below the eigenmode energy, and inversely proportional to the sum of the group-velocity magnitudes of these two eigenmodes.

This flux is computed in momentum space: on the selected band it is taken over the range of $k$ where the dispersion averaged over $p$ lies below the eigenmode energy, and on the bands below it is taken over the whole band. This range of $k$ is set by the averaged dispersion because every level of the ladder has this average as the real part of its energy at $k$. Even when the gradient is too weak to mix in other bands, the bands below still affect which two modes make up an eigenmode under OBCs in $x$. When the sum of the Chern numbers of the bands below is zero, the two are the right-moving and left-moving modes on the same level. When this sum is nonzero, the two modes belong to two levels whose indices differ by the sum. Each band below thereby contributes $2\pi$ times its Chern number to the flux. When the selected band itself has zero Chern number, there is still a skin effect as long as the Berry-curvature flux below the eigenmode energy is nonzero. Whether the strip has OBCs in $y$ or a smooth confining potential, the inverse skin length agrees with the formula (Sec.~\ref{sec:display}). When the dispersion is held fixed and only the Berry curvature is changed, the inverse skin length follows the difference of the two Wannier centers and can reverse sign; when this difference does not change, neither does the inverse skin length (Sec.~\ref{sec:source}).

At still weaker gradients, the change of the gain rate across the width of the strip, or across the range set by the confining potential, is no longer much larger than the bandwidth. The ladder then no longer forms within the strip, and the positions of the eigenmodes are set by the smooth confining potential. At fixed $k$, we define the separatrix energy as the maximum over $p$ of the energy that the selected band has at the minimum of the confining potential. For a single-minimum confining potential symmetric about its minimum, the skin effect is suppressed for eigenmodes below the separatrix energy and not for those above it, independent of the Chern number. As the gradient increases, the inverse skin length returns to the formula above (Secs.~\ref{sec:wall} and~\ref{sec:ratio}).

These results show that the noncommutativity of band-projected coordinates gives rise to the noncommutative skin effect. As long as the Berry-curvature flux below the eigenmode energy is nonzero, the bulk bands of an ordinary lattice show this skin effect, and once the ladder forms, the inverse skin length is fixed by this flux. Noncommutative space therefore need not be engineered: a band with Berry curvature is itself a noncommutative space, and a band on which this flux is nonzero is a platform for realizing the noncommutative skin effect. A nonzero flux requires broken time-reversal symmetry. This skin effect can be measured in electric circuits, where grounding elements provide both the gain-loss gradient and the confining potential, and the inverse skin length is obtained from the envelope along $x$ of the OBC eigenmodes (Sec.~\ref{sec:disc}).
\section{The strip, the inverse skin length, and the commutator of projected coordinates}\label{sec:comm}

We first set up the strip and the notation (Fig.~\ref{fig:schematic}). The strip is translation invariant along $x$. We take the lattice constant in $x$ as the unit of length, set $\hbar=1$, and measure energies in units of the nearest-neighbor hopping. The momentum conjugate to $x$ is $k$, with plane waves $\psi_x\propto\beta^x=e^{ikx}$. The strip has a finite width in $y$, and the quasimomentum conjugate to $y$ is $p$. We add two terms that depend on $y$. One is a real confining potential $W(\hat y)$, which depends only on the position operator $\hat y$ and is the same on all orbitals. The other is gain and loss varying linearly in $y$, i.e., the on-site term $ig\tilde y$ with $\tilde y=\hat y-y_0$, where $g$ is the gain-loss gradient; for $g>0$ the side $y>y_0$ has gain and the side $y<y_0$ has loss. The reference point $y_0$ can be chosen freely: changing $y_0$ only adds the same imaginary constant to all energies. Without these two terms the bulk bands are Hermitian, and the selected band is separated from the other bands for all real $p$.

At fixed $k$, the strip reduces to a finite matrix acting on the $y$ and orbital degrees of freedom. With $B_0$ the matrix block within one column of sites and $T$ the hopping block between neighboring columns along $x$, the Bloch matrix of the strip is
\begin{equation}
H_g(\beta)=T\beta^{-1}+B_0+T^\dagger\beta+W(\hat y)+ig\tilde y,
\label{eq:Hg}
\end{equation}
where $\beta=e^{ik}$ and $\hat y$ is the diagonal matrix of the $y$ coordinates of the sites.

We take as our observable the inverse skin length $\kappa$ of the eigenmodes under OBCs in $x$. The envelope of these eigenmodes along $x$ is given by non-Bloch band theory~\cite{Yao2018,Yokomizo2019}. For a complex energy $z$, the determinant in the characteristic equation $\det[z-H_g(\beta)]=0$ is a Laurent polynomial in $\beta$ with lowest and highest powers $-M$ and $M$, so the equation has $2M$ roots. Ordered by $|\beta|$, the two middle roots are the $M$th and the ($M+1$)th, and on the OBC spectrum they satisfy $|\beta_M|=|\beta_{M+1}|$. We define
\begin{equation}
\kappa=-\tfrac12\bigl(\ln|\beta_M|+\ln|\beta_{M+1}|\bigr),
\label{eq:kappa}
\end{equation}
and on the OBC spectrum $\kappa$ is the common $\mathrm{Im}\,k$ of these two roots, with $|\psi_x|\propto e^{-\kappa x}$: for $\kappa>0$ the eigenmodes accumulate at the small-$x$ end, and for $\kappa<0$ at the large-$x$ end.

To find the origin of the skin effect, we start from the coordinate operators. The two real-space coordinate operators commute, $[\hat x,\hat y]=0$. Let $P$ be the projector onto the selected band and $Q=1-P$. The band-projected coordinates $\hat X=P\hat xP$ and $\hat Y=P\hat yP$ in general no longer commute; for a single band their commutator is given by the Berry curvature~\cite[Sec.~VII]{XiaoChangNiu2010}. The commutator follows directly from $[\hat x,\hat y]=0$ and $P+Q=1$. Inserting $1=P+Q$ in the middle of $P\hat x\hat yP$ and of $P\hat y\hat xP$ gives $P\hat x\hat yP=\hat X\hat Y+P\hat xQ\hat yP$ and $P\hat y\hat xP=\hat Y\hat X+P\hat yQ\hat xP$. Since $[\hat x,\hat y]=0$, the two left-hand sides are equal, which gives
\begin{equation}
[\hat X,\hat Y]=P\hat yQ\hat xP-P\hat xQ\hat yP.
\label{eq:comm}
\end{equation}
Both terms on the right pass through intermediate states in the other bands, so the commutator is built from the matrix elements of position between the selected band and the other bands. In the $(k,p)$ representation, the projected coordinates of a single band are $\hat X=i\partial_k+\mathcal A_k$ and $\hat Y=i\partial_p+\mathcal A_p$, where $\mathcal A_k=i\langle u|\partial_ku\rangle$ and $\mathcal A_p=i\langle u|\partial_pu\rangle$ are the Berry connections of the Bloch vector $u(k,p)$ of the selected band; hence
\begin{equation}
[\hat X,\hat Y]=i\Omega(k,p),\qquad \Omega=\partial_k\mathcal A_p-\partial_p\mathcal A_k,
\label{eq:Omega}
\end{equation}
so the commutator is given by the Berry curvature $\Omega$, which has dimension length squared.

What enters the rest of the paper is the part of this commutator averaged over $p$. At fixed $k$, the eigenfunctions of $\hat Y$ are the hybrid Wannier functions arranged along $y$, labeled by an integer $n$. The Wannier spacing $\aW$ is the period of the unit cell along $y$, i.e., the distance between neighboring hybrid Wannier functions, and $p$ runs over the $p$ Brillouin zone (BZ) of length $G=2\pi/\aW$. In the parallel-transport gauge along $p$, $\mathcal A_p$ does not depend on $p$, and $\hat Y=i\partial_p+Y(k)$ takes the value $Y(k)+n\aW$ on the $n$th hybrid Wannier function, where
\begin{equation}
Y(k)=\frac{\aW}{2\pi}\oint_0^{G}\mathcal A_p(k,p)\,dp
\label{eq:Y}
\end{equation}
is the hybrid Wannier center, namely the Berry phase over the $p$ BZ times $\aW/2\pi$~\cite{Taherinejad2014}. A change of gauge shifts $Y(k)$ only by an integer multiple of $\aW$, which amounts to relabeling $n$. The rate at which the Wannier center changes with $k$
\begin{equation}
\theta(k)\equiv\partial_kY=\frac{\aW}{2\pi}\oint_0^G\Omega(k,p)\,dp
\label{eq:theta}
\end{equation}
is the average of the Berry curvature over the $p$ BZ and has dimension length squared. Since $\theta$ is gauge invariant, the second equality can be derived in a smooth gauge periodic in $p$, where $\oint\partial_p\mathcal A_k\,dp=0$. For uniform Berry curvature, $\theta(k)$ is a constant, the $\theta$ of the noncommutative plane in the Introduction. The diagonal matrix element of the commutator, taken in a single hybrid Wannier function, equals $i\theta(k)$: the derivative $i\partial_k$ in $\hat X$ generates translations in $k$, while the value $Y(k)+n\aW$ of $\hat Y$ moves with $k$. Across the BZ along $x$, the Wannier center moves by $\oint\theta\,dk=C\aW$, where $C$ is the Chern number of the band~\cite{Taherinejad2014,Fidkowski2011}.

The gain-loss term carries this commutator into the non-Hermitian spectrum. Since $ig\tilde y$ is linear in $\hat y$, it becomes $ig(\hat Y-y_0)$ after projection onto the selected band: the gain rate a state acquires is set by its projected coordinate $\hat Y$, and the value of $\hat Y$ moves with $k$. As the next section shows, the two modes propagating along $+x$ and along $-x$ at the same energy therefore acquire different gain rates, so that under OBCs the eigenmodes accumulate toward one end in $x$.
\section{Ladder spectrum and inverse skin length under a gain-loss gradient}\label{sec:ladder}

We first consider the bulk problem, unbounded in $y$, without a confining potential. The system is translation invariant along $x$, so $k$ is conserved and the eigenvalue problem in $y$ can be solved at fixed $k$. If a linear real potential is added along $y$, the levels of this problem form a Wannier--Stark ladder~\cite{Wannier1960}, also on topological bands~\cite{Kim2020}, and the positions of the levels are fixed by the Zak phase, that is, by the Wannier center~\cite{Zak1989}. The gain-loss gradient is a linear potential whose real slope is replaced by $ig$. Below we proceed in the same way to find its ladder, which in the single-band case is the imaginary Wannier--Stark ladder~\cite{ZhangChen2023}. As long as the gradient is weak enough not to mix in the neighboring bands, $g|A_{b'b}|\ll\Delta\varepsilon$, we can project the problem onto the selected band. Here $A_{b'b}=i\langle u_{b'}|\partial_pu_b\rangle$ is the interband Berry connection between the selected band $b$ and another band $b'$, and $\Delta\varepsilon$ is the direct gap. Interband coupling corrects the ladder levels at second order in $g$~\cite{DeBeule2024,Combes2016}. The resulting correction to the inverse skin length is two orders higher in $g$ than the leading term, i.e., a relative $O(g^2)$ when the leading term is nonzero (Appendix~\ref{app:ladder}).

The projected wave function arises as follows. At fixed $k$, we expand an eigenstate in the Bloch states $|\psi_{k,p}\rangle$ of the selected band at $(k,p)$, $|\Psi\rangle=\int_0^G dp\,c(p)|\psi_{k,p}\rangle$. The Bloch states are chosen periodic in $p$ with period $G$, so the expansion coefficient $c(p)$ is also periodic in $p$ with period $G$: it is the wave function of this state in $p$ space, the space of the quasimomentum conjugate to $y$. The projected Hamiltonian acts on $c(p)$, and its band-energy term is multiplication by $\varepsilon(k,p)$. Acting on the plane-wave factor $e^{ipy}$, $\hat y$ becomes $i\partial_p$ on $c(p)$ after an integration by parts in $p$. Acting on the $p$-dependent Bloch vector $u(k,p)$, it gives the Berry connection $\mathcal A_p$ after projection. Together, the two parts are exactly $\hat Y=i\partial_p+\mathcal A_p$. The eigenvalue equation is therefore
\begin{equation*}
\bigl[\varepsilon(k,p)+ig\bigl(i\partial_p+\mathcal A_p-y_0\bigr)\bigr]c(p)=z\,c(p),
\end{equation*}
where $\varepsilon(k,p)$ is the dispersion of the selected band in the model without the confining potential and the gain and loss. The derivative term is $ig\cdot i\partial_p=-g\partial_p$, and rearranging gives a first-order linear equation in $p$,
\begin{equation}
\partial_pc=g^{-1}\bigl[\varepsilon(k,p)-z+ig(\mathcal A_p-y_0)\bigr]c,
\label{eq:ode}
\end{equation}
whose solution is
\begin{equation*}
c(p)=c(0)\exp\Bigl\{g^{-1}\!\int_0^p\bigl[\varepsilon-z+ig(\mathcal A_p-y_0)\bigr]dp'\Bigr\},
\end{equation*}
and since $c(p)$ is periodic in $p$ with period $G$, the exponent at $p=G$ must be an integer multiple of $2\pi i$. Writing $\bar\varepsilon(k)=G^{-1}\oint\varepsilon\,dp$ for the $p$-averaged dispersion, using $\oint\mathcal A_p\,dp=GY(k)$ from Eq.~\eqref{eq:Y}, and calling this integer $-n$, we obtain $G[\bar\varepsilon-z+ig(Y-y_0)]=-2\pi ign$, that is,
\begin{equation}
z_n(k)=\bar\varepsilon(k)+ig\bigl[n\aW+Y(k)-y_0\bigr],\;\, n\in\mathbb Z .
\label{eq:ladder}
\end{equation}

\begin{figure}[t]
\includegraphics[width=\columnwidth]{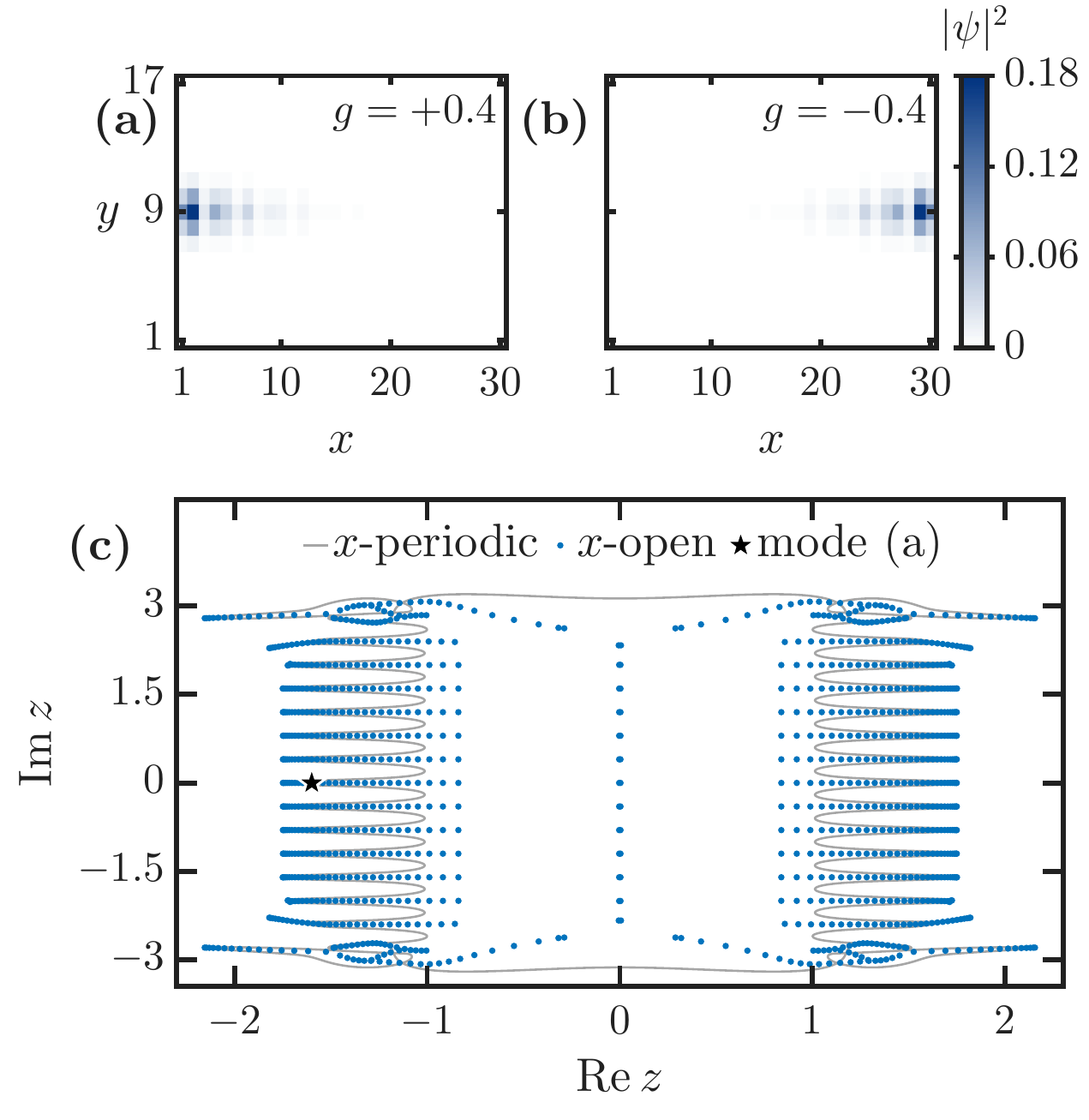}
\caption{In a finite sample, bulk-band eigenmodes accumulate at one end along $x$, and at the other end when $g$ changes sign; apart from a few levels next to the boundaries in $y$, the OBC spectrum forms a ladder equally spaced along the imaginary axis. (a) Intensity $|\psi(x,y)|^2$ of a bulk-band eigenmode in the sample with OBCs: it accumulates at the small-$x$ end and is concentrated in $y$ on a few rows near the center line. (b) The mode with the same eigenvalue after $g$ changes sign, accumulated at the large-$x$ end; its intensity distribution is the point reflection $(x,y)\to(31-x,18-y)$ of (a). (c) Complex spectrum for the same $g$ as in (a). Away from the boundaries in $y$, the OBC eigenvalues (dots) line up on horizontal lines equally spaced along the imaginary axis, which are the levels of the ladder in Eq.~\eqref{eq:ladder}, and they lie inside the closed loops of the spectrum under periodic boundary conditions (PBCs) in $x$ (gray lines). The star marks the mode in (a); the dots in the band gap between the two bands come from chiral edge modes. Parameters: lower band of the Qi--Wu--Zhang (QWZ) model ($m=0.6$, $C=-1$), in a $30\times17$ sample with OBCs in both $x$ and $y$; the PBC spectrum in (c) is computed for the strip with the same $N_y=17$; $|\psi|^2$ is the intensity of the normalized right eigenvector summed over the two orbitals.}
\label{fig:modes}
\end{figure}

Equation~\eqref{eq:ladder} states that the gain-loss gradient turns the $y$ levels at fixed $k$ into a ladder equally spaced along the imaginary axis, with spacing $g\aW$. Within the single-band projection, this result holds for any $g$. All levels share the real part $\bar\varepsilon(k)$, the $p$-averaged dispersion. The imaginary part of the $n$th level, i.e., its gain rate, is the gradient $g$ times the position of its Wannier center $n\aW+Y(k)$ relative to $y_0$. This agrees with the classical intuition that a particle at $y$ acquires the gain rate $g(y-y_0)$, except that ``where the particle is'' is given by the Wannier center. The states of the $n$th level are localized in $y$ near this center. In the OBC spectrum of a finite sample, the bulk-band eigenvalues line up on horizontal lines equally spaced along the imaginary axis, and these lines are the levels [Fig.~\ref{fig:modes}(c)]. Only $\mathcal A_p$ appears in Eq.~\eqref{eq:ode}, and the spectrum depends only on its integral around $p$, $GY(k)$. Since the Hamiltonian contains no potential depending on $\hat x$, $k$ is conserved, and neither $\hat X$ nor $\mathcal A_k$ appears in the eigenvalue problem at fixed $k$. The formula for the skin effect below therefore contains only $Y$ and its derivative $\theta$, the part of the commutator averaged over $p$.

Since the Wannier center $Y(k)$ moves with $k$, states with different $k$ on the same level have different gain rates. We choose $\mathrm{Re}\,z$ such that $\bar\varepsilon(k)=\mathrm{Re}\,z$ has exactly two real roots on the selected band, denoted $k_\pm$. Because all levels have the real part $\bar\varepsilon(k)$, $\bar v_\pm=\partial_k\bar\varepsilon(k_\pm)$ is the group velocity of these two eigenmodes on every level. At $k_+$, $\bar v_+>0$ and the eigenmode propagates along $+x$ (right-moving); at $k_-$, $\bar v_-<0$ and the eigenmode propagates along $-x$ (left-moving). Below we call the magnitude of the group velocity, $|\bar v_\pm|$, the speed. On the same level, the gain rates of the right-moving and left-moving modes differ by $g[Y(k_+)-Y(k_-)]$, the gradient times the difference of the two Wannier centers.

Under OBCs in $x$, the eigenmodes are composed of plane waves with complex wave vectors. Setting $k=k_\pm+iq$ and expanding $z_n$ to first order around $k_\pm$, we have $z_n(k_\pm+iq)\approx z_n(k_\pm)+iq\,[\bar v_\pm+ig\theta(k_\pm)]$. Equating the expansion to $z$ and taking the imaginary part gives the right-moving and left-moving roots of the $n$th level,
\begin{equation}
\mathrm{Im}\,k_\pm(n)=\frac{\mathrm{Im}\,z-g\,[n\aW+Y(k_\pm)-y_0]}{\bar v_\pm}.
\label{eq:lad-lin}
\end{equation}
The linearization holds for $|\kappa\bar\varepsilon''|\ll|\bar v_\pm|$ and $|\kappa\theta'|\ll|\theta|$, i.e., when $k_\pm$ is far enough from the extrema of $\bar\varepsilon$ and $\theta$ varies slowly over a range of wave vectors of size $\kappa$. Equation~\eqref{eq:lad-lin} has a direct meaning. A mode whose gain rate exceeds $\mathrm{Im}\,z$ grows along its own direction of propagation, and a mode whose gain rate is lower decays along it. The spatial growth rate is the difference between the gain rate of the mode and $\mathrm{Im}\,z$, divided by the speed of the mode.

An OBC eigenmode is composed of the plane waves of two roots with equal $|\beta|$. These are the two middle roots of the characteristic equation when its roots are ordered by $|\beta|$ (Sec.~\ref{sec:comm}); one of them is right-moving and the other left-moving. When there is no other band below the selected band, or when the sum of the Chern numbers of the bands below is zero, these two roots come from the same level $n$ of the ladder (Appendix~\ref{app:pair}). Setting $\mathrm{Im}\,k_+(n)=\mathrm{Im}\,k_-(n)$ and using $\bar v_-=-|\bar v_-|$, we find
\begin{equation*}
\mathrm{Im}\,z=g\,(n\aW-y_0)+g\,\frac{|\bar v_-|\,Y(k_+)+|\bar v_+|\,Y(k_-)}{|\bar v_+|+|\bar v_-|},
\end{equation*}
so on the OBC spectrum $\mathrm{Im}\,z$ lies between the gain rates of the two modes. Substituting back into Eq.~\eqref{eq:lad-lin}, the common $\mathrm{Im}\,k$ of the two roots is
\begin{equation}
\kWS=-\frac{g\,[Y(k_+)-Y(k_-)]}{|\bar v_+|+|\bar v_-|}.
\label{eq:kWS}
\end{equation}
The subscript WS stands for Wannier--Stark. By Eq.~\eqref{eq:lad-lin}, one mode then grows along its direction of propagation and the other decays along its own. Since the two propagate in opposite directions, they grow toward the same end at the same rate, and the OBC eigenmode accumulates at this end. For $g[Y(k_+)-Y(k_-)]>0$, the right-moving mode has more gain, $\kWS<0$, and the eigenmode accumulates at the large-$x$ end. When $g$ changes sign, so does $\kWS$, and the eigenmode accumulates at the other end. The inverse skin length $\kWS$ does not depend on the level index $n$: every level of the ladder gives the same inverse skin length. The equal spacing of the ladder and the equal $\kappa$ of all levels hold exactly even without the single-band projection. A translation by one unit cell along $y$ commutes with the bulk Hamiltonian without gain and loss and only adds the constant $ig\aW$ to the gain-loss term, so the set of characteristic roots is invariant under $z\to z+ig\aW$. When the gain and loss vary slowly along $y$, with the on-site term $i\gamma(\hat y)$, $\gamma$ can be replaced, within the $y$ range of one level of the ladder, by its tangent at the position of that level. At leading order, $g$ in Eq.~\eqref{eq:kWS} is then replaced by the local gradient $\gamma'(y_n)$, where $y_n$ is the mean $y$ position of the $n$th level.

The difference of the two Wannier centers is the integral of the commutator, $Y(k_+)-Y(k_-)=\int_{k_-}^{k_+}\theta\,dk$, with the integral taken along the range of $k$ where $\bar\varepsilon<\mathrm{Re}\,z$. The choice of this range, rather than the complementary one, is fixed by which two levels the two middle roots come from (Appendix~\ref{app:pair}). When the Chern number $C$ of the selected band is nonzero, $Y$ increases by $C\aW$ across the BZ, and integrating along the other range would give a result differing by $C\aW$.

\begin{figure*}[t]
\includegraphics[width=\textwidth]{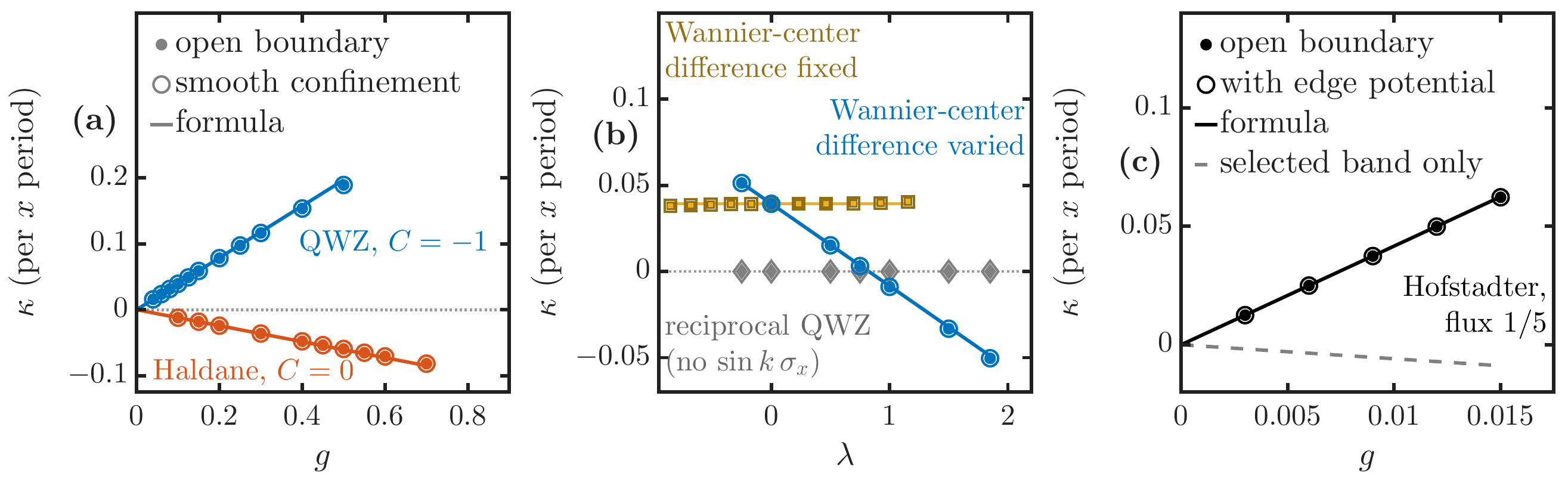}
\caption{In the Wannier--Stark ladder regime, the $\kappa$ of the strip agrees with Eq.~\eqref{eq:kWSgen} and does not depend on the boundary condition in $y$; $\kappa$ is set by the difference of the Wannier centers (the integral of the commutator between $k_-$ and $k_+$) and by the sum of the Chern numbers of the bands below. Symbols are strip numerics and lines are the formula. Filled symbols are for OBCs in $y$. Open symbols are for smooth confinement by a confining potential in (a) and (b), and in (c) for an additional on-site potential of strength $1$ on the outermost row of sites (the edge potential of the legend). (a) $\kappa$ versus $g$. The lower band of the QWZ model (Chern number $-1$, blue) and the conduction band of the Haldane model (zero Chern number, red) both show a skin effect. The sign of each slope is set by the respective difference of Wannier centers and is opposite for these two parameter sets; all points in (a) lie in the ladder regime. (b) $\kappa$ on the QWZ lower band versus the parameter $\lambda$ of the isospectral transformation. For the blue circles, the transformation $U_\lambda=e^{-i\lambda\sin p\,\sigma_x}$ changes the difference of the Wannier centers, and $\kappa$ moves along the straight line $\kWS(\lambda)$ and reverses sign. For the yellow squares, $\sin p$ in the exponent of $U_\lambda$ is replaced by $\sin p+\alpha\sin2p$; this transformation keeps the Wannier centers at $k_\pm$ fixed, and $\kappa$ stays at its value at $\lambda=0$. The gray diamonds are for the reciprocal QWZ model without $\sin k\,\sigma_x$, whose $\kappa$ is zero throughout. (c) $\kappa$ versus $g$ on band $2$ of the Hofstadter model, where the one band below has Chern number $-1$. The solid line is the full Eq.~\eqref{eq:kWSgen}; the dashed line keeps only the term of the selected band itself [Eq.~\eqref{eq:kWS}] and gives a skin-effect direction opposite to that of the strip. Parameters: in all three panels, $y_0$ lies on the center line of the strip; QWZ lower band, $m=0.6$, $\mathrm{Re}\,z=-1.6$; Haldane conduction band, $M_{\mathrm H}=1$, $t_2=0.1$, $\varphi_{\mathrm H}=\pi/2$, $\mathrm{Re}\,z=1.8388$; in (a) and (b), $N_y=31$--$61$ and exponential confining potentials with $\ell=2$ (QWZ) and $1.5$ (Haldane); in (b), $g=0.1$ and $\alpha=1.7414$; in (c), flux per plaquette $1/5$, $\mathrm{Re}\,z=-1.16$, and $N_y=201$. The inverse skin length $\kappa$ is given per $x$ period (one site for QWZ, $\sqrt3$ bond lengths for Haldane, and $5$ sites in (c)).}
\label{fig:kappa}
\end{figure*}

If there are bands below the selected band $b$ and the sum of their Chern numbers, $\sum_{b'<b}C_{b'}$, is nonzero, the two middle roots no longer come from the same level: the right-moving root comes from level $n$ and the left-moving root from level $n'$, with $n'-n=-\sum_{b'<b}C_{b'}$. Under the condition $g|A_{b'b}|\ll\Delta\varepsilon$ stated at the beginning of this section, the bands below do not mix into the selected band at leading order. They affect only which two levels the two middle roots come from. The single-band projection gives a pair of roots on each level of the ladder [Eq.~\eqref{eq:lad-lin}]. When $y$ is unbounded, every level has a pair of roots, there are infinitely many roots, and ``the two middle roots'' are not defined. The two middle roots must be identified on a strip of finite width, among all $2M$ roots of the characteristic equation, and these include roots from the bands below and from the chiral edge modes in the band gap. In the band gap between the selected band and the bands below, each edge carries chiral edge modes. As $k$ runs over one period, the states below the gap move by a net $\sum_{b'<b}C_{b'}$ Wannier spacings toward $+y$ through the chiral edge modes on the two edges. The levels of the two middle roots therefore differ by the same amount. When the eigenmode is far from both edges in $y$, the difference between the levels of the two roots does not depend on the boundary condition in $y$ [Fig.~\ref{fig:kappa}(c) shows this with two boundary conditions]. The full derivation is in Appendix~\ref{app:pair}. Redoing the elimination above with $\mathrm{Im}\,k_+(n)=\mathrm{Im}\,k_-(n')$, the difference $Y(k_+)-Y(k_-)$ in Eq.~\eqref{eq:kWS} acquires an extra $-(n'-n)\aW$, and we obtain
\begin{equation}
\kWS=-\frac{g}{|\bar v_+|+|\bar v_-|}\Bigl[\int_{k_-}^{k_+}\!\theta\,dk+\aW\sum_{b'<b}C_{b'}\Bigr].
\label{eq:kWSgen}
\end{equation}
We denote by $\Phi$ the Berry-curvature flux, in momentum space, of all states below the eigenmode energy,
\begin{equation}
\Phi=\frac{2\pi}{\aW}\Bigl[\int_{k_-}^{k_+}\!\theta\,dk+\aW\sum_{b'<b}C_{b'}\Bigr].
\label{eq:Phi}
\end{equation}
By Eq.~\eqref{eq:theta}, the first term is the flux of the selected band over the range of $k$ where $\bar\varepsilon(k)<\mathrm{Re}\,z$ (for all $p$). The states of the selected band are sorted by $\bar\varepsilon(k)$ because every level of the ladder has $\bar\varepsilon(k)$ as the real part of its energy at $k$. The second term is the sum of the fluxes $2\pi C_{b'}$ of the whole bands below. Equation~\eqref{eq:kWSgen} then takes the form
\begin{equation}
\kWS=-\frac{g\aW}{2\pi}\,\frac{\Phi}{|\bar v_+|+|\bar v_-|}.
\label{eq:kPhi}
\end{equation}
To leading order in the gradient, the inverse skin length is proportional to $g$ and to the flux $\Phi$, and inversely proportional to the sum of the speeds of the two eigenmodes. When $\Phi$ reverses sign, the skin effect reverses direction as well.

On a strip of finite width, the range of gradients in which the ladder and the formula above hold is found by comparing the gradient with the bandwidth. Let $\cW$ be the range of $\varepsilon(k,p)$ over $p$ at fixed $k$, i.e., the bandwidth of this band at that $k$. For the ladder to form, the difference $gL_y$ between the gain rates on the two sides of the strip must be much larger than $\cW$, $gL_y/\cW\gg1$, where $L_y=N_y\aW$ is the width of the strip in $y$ and $N_y$ is the number of unit cells along $y$ (rows for QWZ, chains for Haldane). We call this range the Wannier--Stark ladder regime. On a finite strip, Eq.~\eqref{eq:kWSgen} holds in the ladder regime. The length $\cW/g$ is the distance in $y$ over which the gain rate changes by one bandwidth. When the distance in $y$ from the eigenmode to each edge is much larger than $\cW/g$, the level of the eigenmode is not affected by the boundary condition in $y$. Then $\kappa$ approaches $\kWS$, and OBCs in $y$ and a smooth confining potential give the same value (Sec.~\ref{sec:display}).
\section{Size and direction of the skin effect in the ladder regime, and the boundary condition in \texorpdfstring{$y$}{y}}\label{sec:display}

In this section we look at how large the skin effect given by Eq.~\eqref{eq:kWSgen} is, toward which end it points, whether it depends on the boundary condition in $y$, and how the bands below the selected band contribute to it. The first model is the QWZ model~\cite{QiWuZhang2006}, the simplest two-band Chern insulator, with Bloch Hamiltonian $h_0(k,p)=\bm d\cdot\bm\sigma$, $\bm d=(\sin k,\ \sin p,\ m+\cos k+\cos p)$, and two bands $\pm|\bm d|$. We take the lower band $\varepsilon=-|\bm d|$, with $\aW=1$, the spacing between neighboring rows, and $N_y$ rows along $y$. At $m=0.6$, the lower band has Chern number $C=-1$. There is no other band below the lower band, so Eq.~\eqref{eq:kWSgen} reduces to the term with the difference of the Wannier centers, which is Eq.~\eqref{eq:kWS}. The second model is a zigzag strip of the Haldane model~\cite{Haldane1988}. It has nearest-neighbor hopping $1$, next-nearest-neighbor hopping $t_2$ with phase $\varphi_{\mathrm H}$, and a staggered sublattice mass $M_{\mathrm H}$. The parameters lie in the trivial phase. We take its conduction band. The conduction band and the valence band below it both have zero Chern number, so the Chern-number term vanishes here too, while the Berry curvature of the conduction band is nonzero. Lengths on this strip are measured in bond lengths: $\aW=1.5$ (the spacing between neighboring chains), and the period along $x$ is $\sqrt3$. The inverse skin length $\kappa$ is still given per $x$ period. Along $y$ we take $N_y$ chains, and the two sublattices sit at different $y$ positions (for the definition of $Y$ in this case, see Appendix~\ref{app:ladder}).

We use two boundary conditions in $y$. One is the OBC in $y$: the sample ends at the outermost row of sites, with no sites outside it. The other is a smooth confining potential, for which we take the exponential confining potential $W(y)=W_c\cosh[(y-\yW)/\ell]$, $W_c=2e^{-L_y/2\ell}$, with length scale $\ell$. The minimum of the confining potential, $\yW$, lies on the center line of the strip, and the potential rises to about $1$ at $\yW\pm L_y/2$. Here $L_y=N_y\aW$, so the confined strip has the same width as the strip with OBCs in $y$. On both sides, the sites extend until the confining potential is high enough for the wave functions to be negligible. The $\kappa$ of the strip was computed from the roots of the characteristic equation by Eq.~\eqref{eq:kappa}, which corresponds to a sample with $N_x\to\infty$ sites along $x$.

We first look at the skin effect itself on a finite sample with OBCs in both $x$ and $y$ (Fig.~\ref{fig:modes}). With $y_0$ on the center line of the strip, we call the level whose eigenvalue has the imaginary part closest to zero the central level of the ladder. On the QWZ lower band with $g>0$, a bulk-band eigenmode of the central level accumulates at the small-$x$ end. When $g$ changes sign, the mode with the same eigenvalue accumulates at the other end. The OBC spectrum lies inside the closed loops of the PBC spectrum in $x$, which is how the skin effect shows up in the spectrum~\cite{Yao2018,Okuma2020}.

The size of the skin effect is shown in Fig.~\ref{fig:kappa}(a), which plots the $\kappa$ of the strip ($N_x\to\infty$) versus $g$; every point was taken on the central level of the ladder. In the ladder regime, the $\kappa$ of both models agrees with the straight lines given by Eq.~\eqref{eq:kWSgen}. The inverse skin length $\kappa$ is linear in $g$, and its slope $-[Y(k_+)-Y(k_-)]/(|\bar v_+|+|\bar v_-|)$ is computed entirely from the bulk bands, with a sign set by the difference of the Wannier centers. For these two parameter sets, the slope is positive on the QWZ lower band and negative on the Haldane conduction band, so the two skin effects point in opposite directions. The skin effect on the Haldane conduction band, whose Chern number is zero, shows that a band with zero Chern number also has a skin effect as long as the flux $\Phi$ is nonzero.

In Fig.~\ref{fig:kappa}(a), the values for OBCs in $y$ (filled) and for a smooth confining potential (open) coincide, so $\kappa$ does not depend on the boundary condition in $y$ (for the reason, see the end of Sec.~\ref{sec:ladder}). The one or two levels next to the boundaries in $y$ are affected by the boundaries, and their $\kappa$ is not given by the formula.

When the bands below the selected band have a nonzero sum of Chern numbers, these bands also contribute to the skin effect, through the Chern-number term in Eq.~\eqref{eq:kWSgen}. For Fig.~\ref{fig:kappa}(c) we use a strip of the square lattice in a uniform magnetic field with flux $1/5$ per plaquette (the Hofstadter model~\cite{Hofstadter1976}). The selected band is band $2$, below which lies one band with Chern number $-1$. The $\kappa$ of the strip grows linearly with $g$ and agrees with the full Eq.~\eqref{eq:kWSgen} (solid line). The two boundary conditions, OBCs in $y$ and an additional on-site potential on the outermost row of sites, give the same value. The dashed line is the result of keeping only the term of the selected band itself, i.e., the integral of $\theta$ along the range of $k$ where $\bar\varepsilon<\mathrm{Re}\,z$, which is Eq.~\eqref{eq:kWS}. The dashed line gives a skin-effect direction opposite to that of the strip. The difference between the solid and the dashed line is exactly the Chern-number term of the band below. None of the states of that band lie at the eigenmode energy, yet the band changes which two levels supply the modes that make up the OBC eigenmode (Sec.~\ref{sec:ladder}). In this example, the Chern-number term reverses the direction of the skin effect. What determines the skin effect is therefore the flux $\Phi$, the Berry-curvature flux of all states below that energy, and not only the part on the selected band.
\section{The skin effect comes from the noncommutativity of projected coordinates}\label{sec:source}

Figures~\ref{fig:kappa}(a) and (c) show that Eq.~\eqref{eq:kWSgen} gives the correct size and direction of the skin effect on bands for which the sum of the Chern numbers of the bands below is zero as well as on bands for which it is nonzero. This formula attributes the skin effect to the flux $\Phi$, i.e., to the integral of the commutator (the Berry curvature) over all states below the eigenmode energy. Below we test this attribution with an isospectral transformation and with situations in which $\Phi$ vanishes, and at the end we look at the limit of uniform Berry curvature.

We first need a transformation that changes only the Wannier centers and not the dispersion, and for this purpose we construct one that changes only the Bloch eigenvectors. On the QWZ strip we take the isospectral transformation
\begin{equation*}
h_\lambda=U_\lambda h_0U_\lambda^\dagger,\qquad U_\lambda=e^{-i\lambda\sin p\,\sigma_x},
\end{equation*}
where $h_0=\bm d\cdot\bm\sigma$ is the QWZ Hamiltonian of Sec.~\ref{sec:display} and $\lambda$ is a real parameter. At every $(k,p)$, $U_\lambda$ is a unitary matrix, so $h_\lambda$ and $h_0$ have the same eigenvalues and the dispersion $\varepsilon(k,p)$ is unchanged point by point. Only the Bloch eigenvectors change, $u\to U_\lambda u$. From $U_\lambda^\dagger\partial_pU_\lambda=-i\lambda\cos p\,\sigma_x$, the Berry connection becomes $\mathcal A_p+\lambda\cos p\,\langle u|\sigma_x|u\rangle$, and integrating over $p$ gives
\begin{equation}
\begin{split}
Y_\lambda(k)&=Y_0(k)+\lambda\,Y_1(k),\\
Y_1(k)&=\frac1{2\pi}\oint_0^{2\pi}\cos p\,\langle u|\sigma_x|u\rangle\,dp ,
\end{split}
\label{eq:Ylam}
\end{equation}
and the average of the Berry curvature over $p$ changes accordingly from $\theta_0=\partial_kY_0$ to $\theta_0+\lambda\,\partial_kY_1$. At leading order in the ladder regime, this transformation only shifts the Wannier centers. The ladder problem of $h_\lambda$ turns into that of $h_0+ig\lambda\cos p\,\sigma_x$ when the wave function in the multiband eigenvalue equation in $y$ is written as $U_\lambda$ times a new wave function (Appendix~\ref{app:iso}). The dispersion is unchanged, and the gain-loss term $ig\tilde y$ acquires an additional term $ig\lambda\cos p\,\sigma_x$. Projected onto the selected band, this term is $ig\lambda\cos p\,\langle u|\sigma_x|u\rangle$, whose average over $p$ shifts $Y$ by exactly $\lambda Y_1(k)$. Its interband part only gives $O(g^2)$ corrections. Since $\bar\varepsilon$, $\bar v_\pm$, and $k_\pm$ do not change with $\lambda$, differentiating Eq.~\eqref{eq:kWS} gives
\begin{equation}
\partial_\lambda\kappa=-g\,\frac{\partial_\lambda[Y(k_+)-Y(k_-)]}{|\bar v_+|+|\bar v_-|}=-g\,\frac{Y_1(k_+)-Y_1(k_-)}{|\bar v_+|+|\bar v_-|},
\label{eq:isodk}
\end{equation}
which is independent of $\lambda$: $\kappa$ changes linearly with $\lambda$ and reverses sign at $\lambda_0=-[Y_0(k_+)-Y_0(k_-)]/[Y_1(k_+)-Y_1(k_-)]$, where $Y_\lambda(k_+)-Y_\lambda(k_-)$, computed from the bulk bands, passes through zero.

On the QWZ lower band of Sec.~\ref{sec:display}, we kept the same energy and a fixed $g$ and varied $\lambda$ [Fig.~\ref{fig:kappa}(b)]. The $\kappa$ of the strip moves along the straight line given by Eq.~\eqref{eq:kWS} and reverses sign near the zero of the formula. The values for OBCs in $y$ and for a smooth confining potential coincide. Throughout the scan the dispersion and the speeds stay the same. What changes are the Wannier centers, and the size and direction of the skin effect change with them.

We can also test whether the skin effect is determined by the integral of the commutator between $k_-$ and $k_+$ or by the local value of the Berry curvature. For the transformation $U_\lambda=e^{-i\lambda f(p)\sigma_x}$, the same derivation gives the change of the Wannier center $(\lambda/2\pi)\oint f'(p)\langle u|\sigma_x|u\rangle\,dp$, which for $f=\sin p$ is the $\lambda Y_1(k)$ of Eq.~\eqref{eq:Ylam}. We take $f(p)=\sin p+\alpha\sin2p$ and choose $\alpha$ such that, at the eigenmode energy, the Wannier centers at both $k_+$ and $k_-$ do not change with $\lambda$. The Berry curvature between $k_-$ and $k_+$, and outside this range, then changes with $\lambda$. Yet $Y(k_+)-Y(k_-)$ and the Chern number do not change, and so neither does the $\kappa$ of Eq.~\eqref{eq:kWS}. The yellow squares in Fig.~\ref{fig:kappa}(b) behave exactly this way: the $\kappa$ of the strip stays at its value at $\lambda=0$ (horizontal line), while the blue circles for $f=\sin p$ move along the sloped line and reverse sign. Together, the two data sets show that the skin effect is determined by the integral of the commutator between $k_-$ and $k_+$, not by the local value of the Berry curvature.

Conversely, when the flux $\Phi$ vanishes, this skin effect should disappear. We give two situations in which this happens. The first is time-reversal symmetry, where the Berry curvature can be nonzero but its integral vanishes. We write $H_0(\beta)=T\beta^{-1}+B_0+T^\dagger\beta$ for the strip without the confining potential and the gain and loss. When it has time-reversal symmetry (is reciprocal), $H_0(\beta)^{\mathsf T}=H_0(1/\beta)$ in a real orbital basis. The confining potential and the on-site gain and loss are diagonal matrices and thus invariant under transposition, so $H_g(\beta)^{\mathsf T}=H_g(1/\beta)$. Since a matrix and its transpose have the same determinant, the characteristic roots come in pairs $\beta\leftrightarrow1/\beta$~\cite{KawabataOkumaSato2020}, $|\beta_i||\beta_{2M+1-i}|=1$, the values of $\ln|\beta|$ of the two middle roots sum to zero, and $\kappa=0$. This conclusion holds for any shape of the confining potential, any distribution of gain and loss along $y$, and any $g$. Equation~\eqref{eq:kWSgen} leads to the same conclusion. With time-reversal symmetry, $\Omega(k,p)=-\Omega(-k,-p)$, so $\theta(k)$ is an odd function of $k$, while $\bar\varepsilon(k)$ is an even function and $k_-=-k_+$. Hence $Y(k_+)-Y(k_-)=0$, and the Chern numbers of all bands vanish as well. The gray diamonds in Fig.~\ref{fig:kappa}(b) are for the reciprocal QWZ model without $\sin k\,\sigma_x$; under the same isospectral transformation, its $\kappa$ is zero for all $\lambda$. The second situation does not require time-reversal symmetry. If the commutator vanishes everywhere on the selected band and on the bands below it, i.e., if the Berry curvature of these states vanishes everywhere, then $\theta\equiv0$ and the Chern numbers of these bands are zero. The flux $\Phi$ then vanishes, and the formula gives $\kWS=0$.

Finally, for uniform Berry curvature the formula is not just a leading-order result. Suppose that the Berry curvature is the same everywhere, so that the Wannier center is $Y=\theta k$ with constant $\theta$, and take the dispersion $\bar\varepsilon=k^2$. Equation~\eqref{eq:kWS} then gives $\kappa=-\theta g/2$, independent of the magnitude of $g$ and of the energy. In this limit Eq.~\eqref{eq:ladder} reads $z=k^2+ig(\theta k+n\aW-y_0)$, a quadratic equation in $k$ whose two roots, $k=-i\theta g/2\pm s_n$ with $s_n=\tfrac12\{4[z-ig(n\aW-y_0)]-g^2\theta^2\}^{1/2}$, are symmetric about $\mathrm{Im}\,k=-\theta g/2$. Choose $z$ such that some level $n$ satisfies $\mathrm{Im}\,z=g(n\aW-y_0)$ and $\mathrm{Re}\,z>(g\theta/2)^2$. The $s_n$ of this level is then real, the $\mathrm{Im}\,k$ of both of its roots equals $-\theta g/2$, and their $|\beta|$ are equal. The two roots of every other level lie symmetrically on the two sides of this line, so this pair is precisely the middle pair, and $\kappa=-\theta g/2$ holds exactly for any $g$. This is the skin effect obtained on the noncommutative plane in our earlier work~\cite{WeiKouCPL}. There the commutator of the two coordinate operators is an externally imposed constant $i\theta$, which corresponds to uniform Berry curvature here (the earlier work counts accumulation at the large-$x$ end as positive, which differs from the convention here by a sign). In a lattice, this constant is replaced by the $\theta(k)$ obtained from band projection.
\section{Inverse skin length at weak gradients and the separatrix energy}\label{sec:wall}

So far we have been in the ladder regime. When the change of the gain rate across the width of the strip, or across the range set by the confining potential, is no longer much larger than the bandwidth $\cW$, the $y$ levels are set by the confining potential, and the ladder of Sec.~\ref{sec:ladder} no longer forms. The skin effect of a strip with OBCs in $y$ then also contains a contribution set by the boundaries. At weak gradients, the skin effect is determined by the actual positions of the eigenmodes, and this section and the next discuss it for a smooth confining potential.

\begin{figure*}[t]
\includegraphics[width=\textwidth]{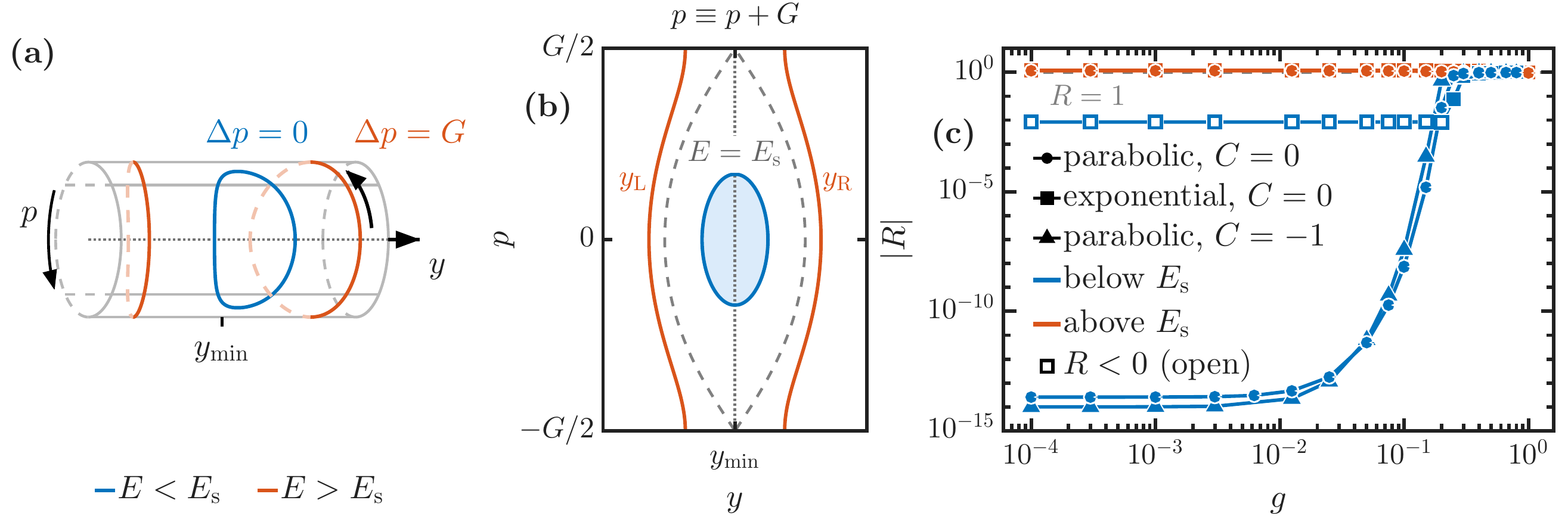}
\caption{For a smooth confining potential with a single minimum, symmetric about it, the skin effect of eigenmodes below the separatrix energy $\Es$ is suppressed at weak gradients and that of eigenmodes above $\Es$ is not, independent of the Chern number of the band. As the gradient increases, the $\kappa$ of both classes of eigenmodes returns to Eq.~\eqref{eq:kWSgen}. (a) The $(y,p)$ phase space at fixed $k$ is a cylinder, with $y$ along its axis and $p$ around it. An orbit below $\Es$ (blue) closes around the minimum $\yW$ of the confining potential without winding around the cylinder, so its net displacement in $p$ is $\Delta p=0$. The two orbits above $\Es$ (red) lie on the two sides of $\yW$, and each winds once around the cylinder, $\Delta p=G$. (b) The same constant-energy contours unrolled onto the $(y,p)$ plane: blue, the closed orbit; red, the two lines $y_{\mathrm L}(p)$ and $y_{\mathrm R}(p)$, each spanning the full $p$ BZ; gray dashed, the separatrix $E=\Es$. (c) $|R|$ versus $g$, where $R$ is the ratio of the $\kappa$ of the strip to the value given by the difference of the Wannier centers [Eq.~\eqref{eq:Rdef}]. The three sets below $\Es$ (blue) are far below $1$ at weak gradients. The two sets above $\Es$ (red), both on the band with zero Chern number, one with a parabolic and one with an exponential confining potential, nearly coincide in the plot and are close to $1$. As $g$ increases, all five sets approach $1$. Symbol shapes distinguish the confining potentials and bands, and open symbols indicate $R<0$. Parameters: (a), (b) QWZ model ($m=3$), $k=0.3$, parabolic confining potential $W=K(y-\yW)^2/2$ with $K=2.66$ chosen for illustration; (c) the three sets below $\Es$ lie on the lowest subband in $y$ and the two sets above $\Es$ on subbands localized on one side of the confining potential, with $k_+$ the right-moving real root at $g=0$; parabolic confining potential, band with zero Chern number: QWZ $m=3$, $K=0.03912$, $k_+=0.470$ (below $\Es$) and $k_+=2.362$ (above $\Es$); symmetric exponential confining potential, band with zero Chern number: QWZ $m=3$, $N_y=7$, $\ell=3$, $k_+=0.510$ (below $\Es$) and $k_+=2.477$ (above $\Es$); parabolic confining potential, band with Chern number $-1$: QWZ $m=1$, $K=0.0335$, $k_+=0.476$ (below $\Es$).}
\label{fig:separatrix}
\end{figure*}

We first expand $\kappa$ in $g$. Take a real energy $E$ and suppose that at $g=0$, $\det[E-H_{g=0}(e^{ik})]=0$ has exactly two roots on the real $k$ axis: the eigenmode at $k_+$ is right-moving and the one at $k_-$ is left-moving, with group velocities $v_+>0>v_-$. The notation follows Sec.~\ref{sec:ladder}, except that $k_\pm$ are now fixed by the dispersion of the strip itself at $g=0$. At $g=0$ the strip is Hermitian for real $k$, and first-order perturbation theory shifts the eigenvalues of these two eigenmodes by $ig\langle\tilde y\rangle_\pm$, where $\langle\tilde y\rangle_\pm$ are their mean $y$ positions (relative to $y_0$). For chiral edge modes, it has been shown that the deviation of the imaginary part of a non-Hermitian perturbation from its average along the closed boundary, divided by the group velocity, gives the local inverse localization length~\cite{Ma2024}. Here we apply this form to the right-moving and left-moving pair of bulk-band eigenmodes at the same energy. As in Sec.~\ref{sec:ladder}, we set the eigenvalue equal to a complex energy $z$ with real part $E$ and linearize around $k_\pm$, which gives $\mathrm{Im}\,k_\pm=(\mathrm{Im}\,z-g\langle\tilde y\rangle_\pm)/v_\pm$. Requiring the two roots to have equal $|\beta|$ fixes $\mathrm{Im}\,z=g(\langle\tilde y\rangle_+|v_-|+\langle\tilde y\rangle_-|v_+|)/(|v_+|+|v_-|)$. Substituting back gives the inverse skin length on the OBC spectrum, to first order in $g$,
\begin{equation}
\kappa_1=-g\,\frac{\langle\tilde y\rangle_+-\langle\tilde y\rangle_-}{|v_+|+|v_-|}
\label{eq:kappa1}
\end{equation}
(Appendix~\ref{app:first} shows that these two roots are precisely the two middle roots ordered by $|\beta|$). First-order perturbation theory holds for $g|\langle\nu|y|\mu\rangle|\ll|E_\nu-E_\mu|$, where $\nu$ and $\mu$ are eigenstates in $y$ at fixed $k$ and $g=0$. Equation~\eqref{eq:kappa1} has the same form as Eq.~\eqref{eq:kWS}, except that the two Wannier centers are replaced by the actual mean $y$ positions of the two eigenmodes, and the speeds by the group-velocity magnitudes of these two eigenmodes on the strip at $g=0$. At weak gradients, therefore, the gradient produces a skin effect along $x$ only when the mean $y$ positions of the two eigenmodes differ.

The question thus becomes: how is the difference of the actual positions of the two eigenmodes related to the difference of the Wannier centers? In the ladder regime, each level is localized near its own Wannier center, and the two differences are the same. At weak gradients, the positions of the eigenmodes are set by the confining potential. We now show that for a smooth confining potential with a single minimum, the answer depends on whether the eigenmode energy lies above or below the separatrix energy.

At fixed $k$, we project onto the selected band and use the parallel-transport gauge along $p$. When the confining potential varies slowly over one Wannier spacing, the projected Hamiltonian in $y$ is, at leading order, $\varepsilon(k,p)+W(\hat Y)$ with $\hat Y=i\partial_p+Y(k)$. For a linear potential this form is exact. Under the same condition a semiclassical picture applies. Since $p$ is a periodic variable, the $(y,p)$ phase space is a cylinder [Fig.~\ref{fig:separatrix}(a)], with $y$ along its axis, and going once around the cylinder in $p$ means running through the $p$ BZ. The classical constant-energy contours are given by $\varepsilon(k,p)+W(y)=E$. We take the confining potential to have a single minimum, at $\yW$, with lowest value $\Wmin$, to be monotonic on each side of $\yW$, and to rise on both sides above $E-\min_p\varepsilon$. We write
\begin{equation}
\Es(k)=\Wmin+\max_p\varepsilon(k,p),
\label{eq:Es}
\end{equation}
the highest energy of the selected band at the minimum of the confining potential, and call it the separatrix energy. Whether $E$ lies below or above $\Es(k)$ determines the shape of the constant-energy contours [Figs.~\ref{fig:separatrix}(a) and (b)]. The classification below applies only to eigenmodes whose energies are far enough from $\Es(k)$. When $E$ is close to $\Es(k)$, the orbits come close to the separatrix, their period diverges, and the semiclassical classification does not apply. A classification by the two sides of the separatrix has a precedent in a lattice with a parabolic confining potential (a harmonically confined lattice). There the momentum space can be mapped onto a quantum pendulum, and the levels fall into two classes above and below the separatrix~\cite{AliHolthausMeier2025}. The states above the barrier are localized at the edges of the trap~\cite{HooleyQuintanilla2004}.

For energies below the separatrix energy, $\Wmin+\min_p\varepsilon<E<\Es(k)$, every $y$ has $E-W(y)\le E-\Wmin<\max_p\varepsilon$, so the values of $p$ at which $\varepsilon(k,p)$ is close to its maximum cannot be reached. The allowed $p$ cover only part of the BZ, and the constant-energy contour closes around the minimum. Along the closed orbit $p$ first increases and then decreases, and after one round it returns to its starting value with zero net displacement in $p$ [the blue orbit in Fig.~\ref{fig:separatrix}(a)]. The Berry phase appears in the Bohr--Sommerfeld quantization condition~\cite[Sec.~VII]{XiaoChangNiu2010}. Here it appears in the form of the Wannier center $Y$. Along an orbit, the Wentzel--Kramers--Brillouin (WKB) wave function is $c(p)\propto\exp[-i\!\int^p(y-Y)\,dp']$, and the phase accumulated once around the orbit is $\oint(y-Y)\,dp=\oint y\,dp-Y\oint dp$, so the coefficient of $Y$ is the net displacement in $p$ over one round, $\oint dp$. On a closed orbit the net displacement is zero, and $Y$ drops out of the quantization condition.

The Wannier center $Y$ has not, however, disappeared from the spectrum altogether. Let $\delta=Y-\yW$ be the displacement of the Wannier center from the minimum of the confining potential, so that $\hat Y-\yW=i\partial_p+\delta$. A shift of the minimum of the confining potential corresponds to a Bloch phase in momentum space~\cite{OzawaPriceCarusotto2015}. The factor $e^{i\aW p}$ is single valued on the $p$ BZ, and a transformation with it shifts $\delta$ by $\aW$, so the spectrum is periodic in $\delta$ with period $\aW$. The spectrum can therefore be Fourier expanded in $e^{iwG\delta}$, with integer harmonic index $w$. Only the harmonics with $w\ne0$ contain $Y$; the zeroth harmonic does not. In the semiclassical picture, the $w$th harmonic corresponds to paths that wind $w$ times around the cylinder. Since the closed orbit itself does not wind around the cylinder, such paths must pass through classically forbidden regions of $p$ space. Their amplitude is $e^{-|w|S}e^{iwG\delta}$, where $S$ is the WKB action under the barrier in $p$ space. To leading order in the WKB approximation, therefore, the Wannier center enters the spectrum of eigenmodes below the separatrix energy only through tunneling, with a weight suppressed as $e^{-S}$.

This is a statement about the spectrum. The skin effect at weak gradients depends on the positions of the modes [Eq.~\eqref{eq:kappa1}], and going from the spectrum to the positions takes one more argument. To obtain the mean $y$ position of a mode, imagine adding a uniform external force along $y$: the derivative of the eigenenergy with respect to this force at zero force gives the position of the mode (Hellmann--Feynman theorem). All terms of the energy that contain $Y$ carry the tunneling factor, which the derivative does not change, so the part of the mode position that contains $Y$ also appears only at order $e^{-S}$. The part of the mean $y$ position of an eigenmode that does not contain $Y$ is set by the closed orbit itself. When the confining potential is symmetric about its minimum, so is the closed orbit, and this part of the mean position lies at the minimum, independent of $k$. For such confining potentials, therefore, at semiclassical leading order there is no $O(1)$ position difference between two modes below the separatrix energy.

For bulk-band eigenmodes with energies above the separatrix energy, $E>\Es(k)$, the situation is entirely different. For every $p$, $E-\varepsilon(k,p)$ exceeds $\Wmin$, and $W(y)=E-\varepsilon(k,p)$ has one solution on each side of the minimum, $y_{\mathrm L}(p)<\yW<y_{\mathrm R}(p)$. Both solutions are single-valued periodic functions of $p$ [the two red lines in Fig.~\ref{fig:separatrix}(b)]. The constant-energy contour thus splits into two orbits, on the left and right sides of the confining potential ($j=\mathrm L,\mathrm R$), each winding once around the cylinder. Along such an orbit the net displacement in $p$ is $G$ [the red orbits in Fig.~\ref{fig:separatrix}(a)], and $Y$ stays in the quantization condition with an $O(1)$ coefficient:
\begin{equation}
\langle y_j\rangle_p=n\aW+Y(k),
\label{eq:quant}
\end{equation}
where $\langle\cdot\rangle_p=G^{-1}\!\int_0^G(\cdot)\,dp$ is the average over $p$ along the orbit and $n$ is an integer (Appendix~\ref{app:wall} gives the WKB derivation). For a linear potential, Eq.~\eqref{eq:quant} is precisely the single-band Wannier--Stark ladder~\cite[Sec.~VII]{XiaoChangNiu2010}. For a smooth confining potential, it states that the $p$-averaged position of a mode falls on the equally spaced positions $n\aW+Y(k)$. A subband in $y$ is a branch of the $y$ levels at fixed $k$, followed continuously as a function of $k$. Consider two modes at the same energy on the same side and with the same $n$, for example the right-moving mode $k_+$ and the left-moving mode $k_-$ on one subband. Subtracting their quantization conditions at leading order shows that the difference of their $p$-averaged positions is $Y(k_+)-Y(k_-)$. When the confining potential has curvature, the mean $y$ position of a mode, which enters Eq.~\eqref{eq:kappa1}, is weighted by the dwell time and differs from the $p$-averaged position by a correction that comes from the curvature. The position difference of the two eigenmodes therefore equals the difference of the Wannier centers plus the difference of these corrections. For a linear potential this correction vanishes (Appendix~\ref{app:wall}).

The division into two classes of orbits uses only the shapes of $\varepsilon$ and $W$, not the Berry connection, so it does not depend on whether the band has a nonzero Chern number. The classification itself is known. What it decides here is whether the Wannier center appears with an $O(1)$ coefficient in the positions of the eigenmodes. At weak gradients, the skin effect is proportional to the position difference of the two eigenmodes [Eq.~\eqref{eq:kappa1}]. When the right-moving and left-moving eigenmodes at the same energy on one subband both lie above the separatrix energy, the difference of the Wannier centers enters their position difference with an $O(1)$ coefficient, and its contribution to the skin effect is not reduced by the tunneling factor that appears below the separatrix energy. When both have energies below the separatrix energy and the confining potential has a single minimum and is symmetric about it, the skin effect is suppressed. The next section compares these two classes of eigenmodes numerically and follows them as the gradient increases.
\section{Skin effect of the two classes of modes under smooth confinement}\label{sec:ratio}

To compare numerically the two classes of eigenmodes above and below the separatrix energy, we use a ratio $R$. At weak gradients, it is the ratio of the difference of the actual mean $y$ positions of the two eigenmodes to the difference of the Wannier centers, and so it tells us directly how much of the position difference of the eigenmodes comes from the Wannier centers. At $g=0$, we take the pair of eigenmodes to be the right-moving and left-moving real roots at the same energy on the same subband in $y$. For $g>0$, we keep $k$ fixed at the $g=0$ value of $k_+$ and follow the eigenvalue $z_+(g)$ of $H_g$ at this $k$ continuously from its value at $g=0$. We read $\kappa(g)$ off the OBC spectrum at $\mathrm{Re}\,z=\mathrm{Re}\,z_+(g)$: it is the common $\mathrm{Im}\,k$ of the pair of eigenmodes continued from $g=0$, and at every $g$ the roots of this pair are the two middle roots. The ratio is defined as
\begin{equation}
R(g)=\frac{\kappa(g)}{\kappa_{\mathrm{ref}}(g)},\qquad
\kappa_{\mathrm{ref}}=-g\,\frac{Y(k_+)-Y(k_-)}{|v_+(g)|+|v_-(g)|},
\label{eq:Rdef}
\end{equation}
i.e., the ratio of the inverse skin length $\kappa$ of the strip to the value $\kappa_{\mathrm{ref}}$ given by the difference of the Wannier centers. The reference value $\kappa_{\mathrm{ref}}$ has the form of Eq.~\eqref{eq:kWS}. In it, $k_\pm$ are the real parts of the two complex wave vectors of this pair, which reduce to the real roots at $g=0$. The velocity $v_\pm(g)$ is the real part of the derivative with respect to $k$ of the branch of eigenvalues of $H_g$ that passes through the corresponding mode, taken at the real $k=k_\pm$, and $|v_\pm(g)|$ is the speed. The difference $Y(k_+)-Y(k_-)$ is the integral of $\theta$ from $k_-$ to $k_+$ through the range of $k$ in which $\bar\varepsilon(k)$ lies below the eigenmode energy. For every set in Fig.~\ref{fig:separatrix}(c), the sum of the Chern numbers of the bands below the selected band is zero, so Eq.~\eqref{eq:kWSgen} is Eq.~\eqref{eq:kWS}. In the ladder regime, the levels of the strip are the levels of the ladder, the group velocities are $\bar v_\pm$, and $\kappa_{\mathrm{ref}}$ is $\kWS$, so $R=1$ corresponds to the formula of Sec.~\ref{sec:ladder}. In the weak-gradient limit, Eq.~\eqref{eq:kappa1} gives the ratio of position differences described above,
\begin{equation*}
R=\frac{\langle\tilde y\rangle_+-\langle\tilde y\rangle_-}{Y(k_+)-Y(k_-)}.
\end{equation*}

Figure~\ref{fig:separatrix}(c) shows $|R|$ versus $g$ for five sets of eigenmodes, with confining potentials symmetric about their minima: the parabolic confining potential $W=K(y-\yW)^2/2$ and the exponential confining potential of Sec.~\ref{sec:display}. At weak gradients the two classes of modes behave entirely differently. For modes below the separatrix energy, the skin effect is strongly suppressed. For the parabolic confining potential, $|R|$ is smaller than $1$ by more than ten orders of magnitude. For the symmetric exponential confining potential, $|R|$ is smaller than $1$ by about two orders of magnitude, and $R$ is negative. For modes above the separatrix energy, $R$ is of order one, and the skin effect is given by the difference of the Wannier centers. Whether the skin effect is suppressed depends on the energy relative to the separatrix energy, not on the Chern number. On the band with zero Chern number, the skin effect of the modes above the separatrix energy is not suppressed, and on the band with Chern number $-1$, the skin effect of the modes below the separatrix energy is suppressed just the same.

As the gradient increases, the $y$ levels become a ladder, the $\kappa$ of the suppressed eigenmodes returns to Eq.~\eqref{eq:kWSgen}, and $R$ rises to near $1$. On the exponential confining potential, $R$ first changes from negative to positive and then rises to near $1$ [Fig.~\ref{fig:separatrix}(c)]. For the eigenmodes above the separatrix energy, the ratio stays of order one in the ladder regime. In this way, for both classes of eigenmodes distinguished by the separatrix energy at weak gradients, $\kappa$ approaches the formula of the ladder regime. When the gradient is weak, the confining potential sets the positions of the modes, and the separatrix energy decides whether the Wannier center appears with an $O(1)$ coefficient in these positions. Once in the ladder regime, the positions of the modes are set by the ladder, and each level is localized near its own Wannier center. The skin effect then returns to the value given by the Berry-curvature flux.
\section{Summary and discussion}\label{sec:disc}

It is known that two coordinates no longer commute after projection onto a band and that their commutator is given by the Berry curvature~\cite{XiaoChangNiu2010}. We have shown that, on a lattice strip whose real-space coordinates commute, a gain-loss gradient along $y$ produces a skin effect along $x$ through the noncommutativity of band-projected coordinates: the noncommutative skin effect. In the ladder regime, the inverse skin length of the strip is proportional to the gradient and to the Berry-curvature flux $\Phi$ of all states below the eigenmode energy, inversely proportional to the sum of the speeds of the two eigenmodes [Eq.~\eqref{eq:kPhi}], and independent of the boundary condition in $y$. A band with zero Chern number also has a skin effect as long as $\Phi$ is nonzero. At weak gradients, for a confining potential with a single minimum, symmetric about it, the skin effect of eigenmodes below the separatrix energy is suppressed, while that of eigenmodes above the separatrix energy is not.

For uniform Berry curvature and the dispersion $k^2$, Eq.~\eqref{eq:kWS} holds exactly for any gradient and reduces to the result of our earlier work on the noncommutative plane~\cite{WeiKouCPL}. The noncommutative plane of that work is precisely the special case of uniform Berry curvature. The formula itself does not require uniform Berry curvature, so any band with nonzero flux $\Phi$ (which requires broken time-reversal symmetry) is a platform for realizing the noncommutative skin effect. The lattice brings in three features that are absent when the commutator is a constant. The commutator varies with momentum, and what determines the skin effect is its integral over a range of momenta. Each band below contributes $2\pi$ times its Chern number to the flux. With a confining potential, there is in addition the separatrix energy.

This skin effect can be measured in electric circuits. The on-site gain and loss and the confining potential are both spatial distributions of grounding elements. The admittance matrix of the circuit is $\mathcal L=i\mathcal G_0(H-E)$~\cite{Lee2018topolectrical}, where $H$ is the tight-binding Hamiltonian of the sample, $E$ the chosen real energy, and $\mathcal G_0$ a constant admittance. A circuit with gain can be stabilized by adding a sufficiently large uniform grounding admittance $\mathcal G_e$, $\mathcal L_s=\mathcal L+\mathcal G_e=i\mathcal G_0[H-(E+i\gamma_e)]$ with $\gamma_e=\mathcal G_e/\mathcal G_0$. The uniform grounding only shifts the reference energy by a purely imaginary constant $i\gamma_e$ and leaves the eigenvectors unchanged. Injecting currents node by node and measuring the voltages gives the impedance matrix, which differs from the Green's function of $H$ at the complex energy $E+i\gamma_e$ only by a constant factor~\cite{helbig_GeneralizedBulkBoundary_2020}. Measuring the whole matrix and diagonalizing it yields the OBC eigenmodes, and $\kappa$ is obtained from the envelope of the right eigenmodes along $x$. Single-frequency two-terminal transmission does not give the $\kappa$ of the OBC spectrum, which is why the whole impedance matrix has to be measured.

This measurement differs from existing measurements of band geometry. In a linear real potential, band geometry can already be extracted from time evolution: Bloch oscillations combined with interferometry have measured the Zak phase~\cite{Atala2013}, and the Hall drift has given the Chern number~\cite{Aidelsburger2015}, as proposed in Ref.~\cite{DauphinGoldman2013}. Our observable, in contrast, is a stationary quantity, the inverse skin length of the OBC eigenmodes. The Wannier--Stark ladder spectrum in a linear real potential also reflects topology: the Chern number shows up directly in the winding number of the ladder~\cite{LeePark2015}, so this approach yields an integer. What we obtain in the ladder regime from the inverse skin length and the speeds of the two eigenmodes is the Berry-curvature flux of all states below the eigenmode energy.

An open question concerns bulk bands that are themselves non-Hermitian. Our Wannier centers come from the Berry phase of Hermitian Bloch functions over the $p$ BZ. When the bulk bands are non-Hermitian, parallel transport along $p$ can use the right eigenstates alone or the biorthogonal inner product of left and right eigenstates, and the two choices generally give different Wannier centers. For the static non-Hermitian skin effect, the localization length is known to appear in the quantum metric of the right eigenstates and not in the biorthogonal metric~\cite{ImuraKawabata2026}. Extending our results to systems whose bulk bands themselves have point-gap topology first requires answering which inner product defines the Wannier center that corresponds to the inverse skin length of a confined strip.

We finally return to the question raised in the Introduction. In real systems whose real-space coordinates commute, projection onto a band is a known route to noncommuting coordinates. Through it, the skin effect of noncommutative space arises in the bulk bands of an ordinary lattice, and in the ladder regime its strength is fixed by the Berry-curvature flux below the eigenmode energy. A band with Berry curvature is itself a noncommutative space. In Hermitian systems this noncommutativity brings about the anomalous velocity and the Hall response~\cite{XiaoChangNiu2010}, and in non-Hermitian systems, gain and loss along one direction reveal it as a skin effect along the other.

\begin{acknowledgments}
This work was supported by the National Key R\&D Program of China (Grant No.~2023YFA1406704).
\end{acknowledgments}

\appendix

\section{Single-band projection: orbital positions and interband corrections}\label{app:ladder}
When the orbitals sit at different $y$ positions within the unit cell, we use a gauge that is periodic in $p$ with period $G$ and whose Bloch phase contains only the position of the unit cell, not the positions of the orbitals within it (below, the $G$-periodic gauge). In this gauge, $\hat y=i\partial_p+\hat\tau$, with $\hat\tau=\mathrm{diag}(\tau_\alpha)$ the $y$ positions of the orbitals within the unit cell ($\hat\tau=0$ when all orbitals sit on the lattice sites). Projection onto the selected band still gives $\hat Y=i\partial_p+\mathcal A_p(k,p)$, except that $\mathcal A_p$ and $Y$ now contain the sublattice positions $\langle\tilde u|\hat\tau|\tilde u\rangle$ ($\tilde u$ is the Bloch vector in the $G$-periodic gauge). We still have $\oint\mathcal A_p\,dp=GY$, the eigenvalue equation is still Eq.~\eqref{eq:ode}, and the definition $\theta=\partial_kY$ is unchanged.

Without the single-band projection, the periodicity condition is $e^{G(z+igy_0)/g}\in\mathrm{spec}\,U_\tau(k)$, where $U_\tau=\mathrm P\exp[g^{-1}\!\int_0^G(h^{\mathrm{per}}+ig\hat\tau)\,dp]$ is path ordered along $p$ and $h^{\mathrm{per}}(k,p)$ is the Bloch matrix in the $G$-periodic gauge. This is the Floquet--Bloch condition in a linear real potential $Fy$ (with slope $F$)~\cite{DeBeule2024}, continued by $F\to ig$. Expanding in $g$ gives $z_n=\bar\varepsilon+ig[n\aW+Y-y_0]+g^2\bar\mu_2+g^3\bar\mu_3+O(g^4)$, where $\bar\mu_2=\langle\sum_{b'\ne b}|A_{b'b}|^2/(\varepsilon_{b'}-\varepsilon)\rangle_p$ is real and $\bar\mu_3$ is purely imaginary. In a linear real potential, they correspond to the second-order Stark shift~\cite{DeBeule2024,Combes2016} and to the second-order position response~\cite{Xiang2023}, respectively. The expansion has the same form as Eq.~\eqref{eq:ladder}, with $\bar\varepsilon\to\bar\varepsilon+g^2\bar\mu_2$ and $Y\to Y+g^2\,\mathrm{Im}\,\bar\mu_3$, so the arguments of Sec.~\ref{sec:ladder} on the pairing and on equal $|\beta|$ apply unchanged. Both $\bar v_\pm$ and $Y(k_+)-Y(k_-)$ change only at $O(g^2)$, so the correction to $\kappa$ from interband mixing is $O(g^3)$, with expansion parameter $g|A_{b'b}|/\Delta\varepsilon$. The next term of the linearization in complex $k$ is also $O(g^3)$. Altogether, $\kappa=\kWS+O(g^3)$: the correction is two orders higher in $g$ than the leading order, i.e., a relative $O(g^2)$ when $\kWS$ is nonzero.

\section{Pairing of the two middle roots under OBCs}\label{app:pair}
Here we determine the level of the ladder to which each of the two middle roots of Sec.~\ref{sec:ladder} belongs. We fix the complex energy $z$ and take $g>0$ (for $g<0$, the two sides in $y$ are interchanged). Level $l$ of the ladder gives one right-moving and one left-moving root, whose $\mathrm{Im}\,k$ is given by Eq.~\eqref{eq:lad-lin}: the $\mathrm{Im}\,k_+(l)$ of the right-moving root decreases as $l$ increases, and the $\mathrm{Im}\,k_-(l)$ of the left-moving root increases as $l$ increases. Suppose that the two middle roots are the right-moving root of level $n$ and the left-moving root of level $n'$. Eliminating $\mathrm{Im}\,z$ with $\mathrm{Im}\,k_+(n)=\mathrm{Im}\,k_-(n')$ gives
\begin{equation}
\kappa_{nn'}=-g\,\frac{Y(k_+)-Y(k_-)}{|\bar v_+|+|\bar v_-|}+(n'-n)\,\frac{g\aW}{|\bar v_+|+|\bar v_-|}.
\label{app:kjm}
\end{equation}
What remains is to count $n'-n$. We proceed in four parts.

\emph{First: the middle roots as a winding number.} For real $q$, we apply the argument principle to $\det[z-H_g(\beta)]$ on the circle $|\beta|=e^{-q}$. The number of roots inside the circle, i.e., of roots with $\mathrm{Im}\,k>q$, equals the order $M$ of the pole at $\beta=0$ plus the number of times the determinant winds around zero along the circle. The determinant is $\prod_\nu[z-z_\nu(\beta)]$, where $z_\nu(\beta)$ are the eigenvalues of $H_g(\beta)$ and $\nu$ runs over all $y$ levels (the ladder levels of all bands and the chiral edge modes). This winding number therefore equals the total winding number $w_q(z)$ around $z$ of the family of curves $k\mapsto z_\nu(k+iq)$, with $k$ running over one period. There are $2M$ finite roots. The two middle roots are the $M$th and ($M+1$)th in ascending order of $\mathrm{Im}\,k$, and their common $\mathrm{Im}\,k$ is $\kappa$. For $q$ slightly below $\kappa$, there are $M+1$ roots with $\mathrm{Im}\,k>q$, and $w_q=+1$. For $q$ slightly above $\kappa$, there are $M-1$, and $w_q=-1$. In other words, as $q$ passes $\kappa$, the curves of the two middle roots sweep exactly across $z$. With these two roots removed, the remaining curves wind around $z$ zero times at $q=\kappa$.

\emph{Second: counting the winding number by crossings.} Draw a vertical ray upward from $z$. A point at which the real part of a curve $z_\nu(k+iq)$ passes through $\mathrm{Re}\,z$ is called a crossing. It is a right-moving crossing if the real part increases with $k$ there, and a left-moving crossing if it decreases. If a crossing lies above $z$, the curve crosses the ray there: a right-moving crossing passes from left to right, which is clockwise and counts $-1$, while a left-moving crossing is counterclockwise and counts $+1$. Thus $w_q(z)$ equals the number of left-moving crossings above minus the number of right-moving crossings above. To first order in $q$, the imaginary part at a crossing is $g$ times the position of the state relative to $y_0$, plus $q$ times the signed velocity. This is where Eq.~\eqref{eq:lad-lin} comes from. For level $l$ of the ladder, Eq.~\eqref{eq:lad-lin} states that the right-moving crossing lies above if and only if $\mathrm{Im}\,k_+(l)<q$, and the left-moving crossing lies above if and only if $\mathrm{Im}\,k_-(l)>q$. Taking $q=\kappa=\mathrm{Im}\,k_+(n)=\mathrm{Im}\,k_-(n')$, the monotonicity of the two families of roots gives
\begin{gather*}
\text{right-moving crossing of level $l$ above}\ \Leftrightarrow\ l>n,\\
\text{left-moving crossing of level $l$ above}\ \Leftrightarrow\ l>n'.
\end{gather*}
The result of the first part then becomes: apart from the two middle crossings, there are as many left-moving crossings above as right-moving crossings above.

\emph{Third: pairing the crossings along arcs.} Follow each curve in the direction of increasing $k$. The parts on which its real part lies below $\mathrm{Re}\,z$ form a number of arcs. Each arc begins at a left-moving crossing (where the real part drops below $\mathrm{Re}\,z$) and ends at a right-moving crossing (where it rises back above $\mathrm{Re}\,z$), so all crossings are paired by the arcs. The count ``left-moving crossings above minus right-moving crossings above'' can be summed arc by arc: an arc whose start lies above and whose end does not contributes $+1$; an arc whose end lies above and whose start does not contributes $-1$; an arc whose two ends are both above or both not above contributes $0$. There are two kinds of arcs.

Arcs of the first kind stay in the selected band $b$. Such an arc starts at $k_-$ and runs through the range of $k$ where $\bar\varepsilon<\mathrm{Re}\,z$ to $k_+$. Its two ends are the left-moving and right-moving crossings of the same level $l$, and its contribution is $[l>n']-[l>n]$ (a bracket is $1$ if the statement inside is true and $0$ otherwise; for the two middle crossings, both brackets are exactly $0$). If $n'=n$, every such arc contributes zero. If $n'<n$, the $n-n'$ arcs of levels $n'+1$ through $n$ start above and end not above, each contributing $+1$. If $n'>n$, the arcs of levels $n+1$ through $n'$ each contribute $-1$. In both cases the arcs of the first kind add up to $n-n'$.

Arcs of the second kind pass through the bands below. Such an arc starts in band $b$ near one edge and descends into the bands below through the chiral edge modes in the band gap on that edge. As $k$ increases, it crosses the strip along $y$ within the bands below. It then returns to band $b$ through the chiral edge modes on the other edge, so that its two ends lie at the two edges. The eigenmode is far from both edges in $y$. On the large-$y$ edge the gain is large, the imaginary parts at the crossings there lie far above $\mathrm{Im}\,z$, and all these crossings are above. None of the crossings on the small-$y$ edge is above. Hence an arc from the small-$y$ edge to the large-$y$ edge (start not above, end above) contributes $-1$, and an arc in the opposite direction contributes $+1$. Each time $k$ runs over one period, the Wannier center of each band below moves by $C_{b'}\aW$ toward $+y$ (Sec.~\ref{sec:comm}), and the states below the gap together move by a net $\sum_{b'<b}C_{b'}$ Wannier spacings toward $+y$. These are exactly the states that the arcs of the second kind carry from the small-$y$ edge to the large-$y$ edge. Equivalently, the net number of chiral edge modes on each edge in the band gap equals $\sum_{b'<b}C_{b'}$~\cite{ElbauGraf2002}, which in a linear real potential corresponds to the winding number of the Wannier--Stark ladder being equal to the Chern number~\cite{LeePark2015}. Among the arcs of the second kind, therefore, those from the small-$y$ edge to the large-$y$ edge outnumber those in the opposite direction by $\sum_{b'<b}C_{b'}$, and together they contribute $-\sum_{b'<b}C_{b'}$.

\emph{Fourth: combining the two kinds of arcs.} The balance condition at the end of the second part requires the contributions of the two kinds of arcs to sum to zero, $(n-n')-\sum_{b'<b}C_{b'}=0$, that is,
\begin{equation*}
n'-n=-\sum_{b'<b}C_{b'}.
\end{equation*}
Substituting this into Eq.~\eqref{app:kjm} gives Eq.~\eqref{eq:kWSgen}. For example, when the Chern numbers of the bands below sum to $1$, $n'=n-1$, and the middle pair consists of the right-moving root of level $n$ and the left-moving root of level $n-1$. The arc of the first kind belonging to level $n$ then starts above and ends not above, contributing $+1$. This exactly cancels the $-1$ of one arc of the second kind running from the small-$y$ edge to the large-$y$ edge. When there is no other band below the selected band, or when the Chern numbers of the bands below sum to zero, there is no net arc of the second kind, $n'=n$, and we obtain Eq.~\eqref{eq:kWS}. The counting uses the conditions of Sec.~\ref{sec:ladder}: the eigenmode is far from both edges in $y$, the real parts of the other bands are separated from $\mathrm{Re}\,z$ by a band gap, and $g|A_{b'b}|$ is much smaller than the gap.

\section{Ladder problem under the isospectral transformation}\label{app:iso}
The transformation $U_\lambda=e^{-i\lambda\sin p\,\sigma_x}$ satisfies $U_\lambda(0)=U_\lambda(2\pi)=1$ and $U_\lambda^\dagger(i\partial_p)U_\lambda=i\partial_p+\lambda\cos p\,\sigma_x$. Both orbitals of the QWZ model sit on the lattice sites, and the multiband eigenvalue equation in $y$ at fixed $k$, $[h_\lambda+ig(i\partial_p-y_0)]\chi=z\chi$, is equivalent to $\partial_p\chi=g^{-1}[h_\lambda-z-igy_0]\chi$. Setting $\chi=U_\lambda\tilde\chi$ gives
\begin{equation*}
\partial_p\tilde\chi=g^{-1}\bigl[h_0+ig\lambda\cos p\,\sigma_x-z-igy_0\bigr]\tilde\chi,
\end{equation*}
and since $U_\lambda$ is periodic in $p$, the two problems have the same periodicity condition. The ladder problem of $h_\lambda$ under the gain-loss gradient, unbounded in $y$, is therefore the same as the ladder problem of $h_0+ig\lambda\cos p\,\sigma_x$. At $g=0$ the latter reduces to $h_0$, so its dispersion is that of $h_0$. Projected onto the selected band, its additional term is $ig\lambda\cos p\,\langle u|\sigma_x|u\rangle$, which enters the periodicity condition alongside $ig\mathcal A_p$, and its integral once around $p$ gives Eq.~\eqref{eq:Ylam}. By Appendix~\ref{app:ladder}, the interband part of the additional term gives $O(g^2)$ corrections.

\section{Inverse skin length at weak gradients}\label{app:first}
Let the real energy $E$ lie in an energy interval with only one pair $k_\pm$. At $g=0$, $H_{g=0}(\beta)^\dagger=H_{g=0}(1/\beta^*)$, so the set of roots of the characteristic equation is closed under $\beta\to1/\beta^*$. The roots on the unit circle are the roots at real $k$, which in this interval are exactly $\beta_\pm=e^{ik_\pm}$. The other roots come in pairs, one inside and one outside the circle, so $\beta_\pm$ are the middle roots in the ordering by $|\beta|$. For small $g$, the other roots move by less than their distances to the unit circle, and the ordering is unchanged. At $k_\pm$, $H_{g=0}$ is Hermitian and its left and right eigenvectors coincide. The eigenvalues of the eigenmodes are $E_\pm(k)+ig\langle\tilde y\rangle_\pm+O(g^2)$, where $E_\pm(k)$ are the branches of eigenvalues of $H_{g=0}(k)$ near $k_\pm$ and $v_\pm=\partial_kE_\pm|_{k_\pm}$. Following the derivation in Sec.~\ref{sec:wall}, we then obtain $\mathrm{Im}\,z$ on the OBC spectrum and Eq.~\eqref{eq:kappa1}.

\section{Quantization condition and position difference}\label{app:wall}
Fix a real $k$. When the confining potential varies slowly on the scale of $\aW$, the projected Hamiltonian in $y$ is, at leading order, $H_P=\varepsilon(k,p)+W(\hat Y)$ with $\hat Y=i\partial_p+Y(k)$. For a linear potential this form is exact. Suppose that the selected band is isolated, that the confining potential has a single minimum, that $E>\Es(k)$ and is not close to $\Es(k)$, and that the confining potential exceeds $E-\min_p\varepsilon$ on both sides. Then the two solutions $y_{\mathrm R}(p)>\yW>y_{\mathrm L}(p)$ of Sec.~\ref{sec:wall} exist over the whole $p$ BZ, with $W'(y_j)\ne0$. Setting $c(p)=\exp[-i\!\int^p(\phi-Y)\,dp']$, we have $\hat Yc=\phi\,c$ and $\hat Y^2c=(\phi^2+i\phi')c$, and the eigenvalue equation becomes
\begin{equation}
W(\phi)+\tfrac i2W''(\phi)\,\phi'+\cdots+\varepsilon(k,p)=E.
\label{app:wkb}
\end{equation}
We expand $\phi=\phi_0+\phi_1+\cdots$ in powers of $1/\ell$, where $\ell$ is the length scale of the confining potential. At zeroth order, $W(\phi_0)=E-\varepsilon$, i.e., $\phi_0=y_j(p)$. At first order, $\phi_1=-\tfrac i2\partial_p\ln|W'(y_j)|$ is the total derivative of a periodic function, so $\oint\phi_1\,dp=0$, and it only gives the amplitude $|W'|^{-1/2}$. Because $W'\ne0$, a single WKB solution covers the whole $p$ BZ, with no connection at turning points and no Maslov phase. The periodicity condition $c(p+G)=c(p)$ requires $\oint(\phi-Y)\,dp\in2\pi\mathbb Z$, i.e., $\int_0^G(y_j-Y)\,dp=2\pi n$, and dividing by $G$ gives Eq.~\eqref{eq:quant}. For a linear potential $W=Fy$, it gives $E=\bar\varepsilon+F(n\aW+Y)$, the single-band Wannier--Stark ladder.

Two modes above the separatrix energy on the same side, with the same $n$ and the same energy $E$ ($k_1\ne k_2$), both satisfy Eq.~\eqref{eq:quant}, and subtracting the two conditions gives $\langle y_j\rangle_p(k_1)-\langle y_j\rangle_p(k_2)=Y(k_1)-Y(k_2)$. The amplitude $|W'|^{-1/2}$ makes $|c|^2\propto1/|W'|$, so the mean $y$ position of a mode is the mean position weighted by the dwell time, $\langle y_j/|W'|\rangle_p/\langle1/|W'|\rangle_p=\langle y_j\rangle_p+\zeta_j$, where $\zeta_j=[\langle y_j/|W'|\rangle_p-\langle y_j\rangle_p\langle1/|W'|\rangle_p]/\langle1/|W'|\rangle_p$ is the covariance of $y_j$ and the dwell weight along the orbit, divided by $\langle1/|W'|\rangle_p$. The difference of the weighted positions is therefore $[Y(k_1)-Y(k_2)]+[\zeta_j(k_1)-\zeta_j(k_2)]$, where the correction comes from the curvature of the confining potential. For a linear potential, $W'$ is constant and $\zeta_j=0$.

\end{document}